\documentstyle[prl,aps,multicol,epsf]{revtex}

\begin{document}
\draft

\title{Critical Behaviour of Random Bond Potts Models:
       A Transfer Matrix Study}

\author{Jesper Lykke Jacobsen$^{1,2,3}$ and John Cardy$^{1,4}$}

\address{$^1$University of Oxford, Department of Physics---Theoretical Physics,
             1 Keble Road, Oxford OX1 3NP, U.K. \\
         $^2$Institute of Physics and Astronomy, University of Aarhus,
             Ny Munkegade, DK-8000 Aarhus C, Denmark. \\
         $^3$Somerville College, Oxford; $^4$All Souls College, Oxford.}

\date{\today}

\maketitle

\begin{abstract}

We study the two-dimensional Potts model on the square lattice in the
presence of quenched random-bond impurities. For $q>4$ the first-order
transitions of the pure model are softened due to the impurities, and we
determine the resulting universality classes by combining transfer matrix
data with conformal invariance. The magnetic exponent $\beta/\nu$ varies
continuously with $q$, assuming non-Ising values for $q>4$, whereas the
correlation length exponent $\nu$ is numerically consistent with unity.
We present evidence for the correctness of a formerly proposed phase
diagram, unifying pure, percolative and non-trivial random behaviour.

\end{abstract}



\begin{multicols}{2}

\section{Introduction}

The effect of quenched bond randomness on a classical statistical mechanics
system whose pure version undergoes a second-order phase transition is well
understood. Namely, the so-called Harris criterion states that if
the critical exponent $\alpha^{\rm pure}$ governing the divergence
of the specific heat at the transition point of the {\em pure} system is
negative, weak bond randomness is irrelevant in the renormalisation group
(RG) sense and the pure fixed point (FP) is stable \cite{Harris}. On the
other hand, if $\alpha^{\rm pure} > 0$ the randomness is relevant
and causes a cross-over to critical behaviour governed by a new random FP
nearby, at least if the cross-over exponent $\alpha^{\rm pure}$ is
small.

This should be contrasted with the more dramatical effects of randomness in
the field conjugate to the local magnetisation. Such randomness can
eliminate low-dimensional phase transitions altogether, and at least it
always changes the values of the critical exponents \cite{Aharony}. For
this reason most early research was concentrated on field randomness. In
this context a particularly popular model is the random field Ising model
(RFIM) for which a classical argument due to Imry and Ma \cite{Imry-Ma}
from a simple comparison of the field fluctuations with the stabilising
effect caused by the formation of a domain wall concluded, that the lower
critical dimension of the RFIM is $d_{\rm l} = 2$.

The issue of quenched bond randomness imposed on a system that undergoes a
thermal {\em first}-order phase transition is less studied. An adaptation
of the Imry-Ma argument can be established by noting that the bond
randomness couples to the local energy density, which differs for the two
phases that co-exist at the critical point of the pure system, in exactly
the same way that the random field couples to the local magnetisation in
the RFIM. Consequently the existence of a non-vanishing latent heat for
$d < 2$ can be ruled out.
Early work by Imry and Wortis \cite{Imry-Wortis} furnished a
heuristic argument, reminiscent of that of the Harris criterion, that the
bond randomness indeed softens any such phase transition in $d=2$ to a
continuous one. A subsequent phenomenological RG
argument by Hui and Berker \cite{Hui-Berker} confirmed that the lower
critical dimension for random bond tricriticality and end-point criticality
is $d_{\rm l} = 2$. As the dimensionality increases, tricritical points and
critical end points emerge from $T=0$. Finally, a mathematically rigorous
theorem by Aizenman and Wehr \cite{Aizenman-Wehr} stated that quite generally
for $d \le 2$ an {\em arbitrarily} weak amount of quenched bond randomness
leads to the elimination of any discontinuity in the density of the
variable conjugate to the fluctuating parameter.

The question then emerges whether this softening of the phase
transition can be verified for specific models and, if so, what are the
universality classes of these novel second-order phase transitions.
An investigation along these lines has recently been initiated by one of us
\cite{Cardy96}, by considering a system of $N$ two-dimensional
Ising models coupled by their energy operators which, according to
mean-field theory (MFT), is supposed to display a second-order phase
transition. For $N>2$, however, the RG flow of the model exhibits a runaway
behaviour, which is characteristic of a fluctuation driven first-order
transition \cite{Cardy-book}.
In this sense the transition is only weakly first-order and hence amenable
to perturbative calculations. On adding weak bond randomness it was found
that the RG trajectories curl back towards the pure decoupled Ising FP, and
consequently Ising exponents are expected, up to possible logarithmic
corrections. This study was extended by Pujol \cite{Pujol96} to the case of
$N$ coupled $q$-state random bond Potts models for $2 \le q \le 4$, but
here the universality class of the impurity softened transition was found
to depend on the coupling between the models.

A more interesting model for studying the effect of quenched bond
impurities on a first-order transition is the $q$-state random bond Potts
model (RBPM). For $q>4$ the phase transition of the pure system is first order
with a latent heat that is an increasing function of $q$ \cite{Baxter73}.
In fact, since the transition is first-order already in MFT, on the
RG level it is controlled by a zero-temperature discontinuity FP
with the eigenvalue of the relevant scaling operator being $y=d$
\cite{Cardy-book}. Quenched randomness coupling to the local energy density
thus has the eigenvalue $d-2(d-y)=d$ and is strongly relevant, whence an RG
treatment appears to be problematic.

The work undertaken until now has therefore mainly been
numerical. Extensive Monte Carlo (MC) simulations have been carried out for
$q=8$ by Chen, Ferrenberg and Landau \cite{Chen95} confirming the
transition softening scenario outlined above, and finding critical exponents
numerically consistent with those of the {\em pure} Ising model. Similar
conclusions were reached by Domany and Wiseman \cite{Domany95} for
$q=4$ and also for the Ashkin-Teller model. It thus appears that in a
variety of situations the universality class of the bond disordered models
is that of the Ising model, {\em irrespective} of the symmetry underlying the
original model.

To explain these findings Kardar {\em et al.}~\cite{Kardar95} have proposed
an interface model for the RBPM which, after several approximations, is
amenable to an RG treatment that is exact on the hierarchical lattice. In
the pure model the interface exhibits a branching structure with fractal
dimension at criticality, but when randomness is present the critical
interface is asymptotically linear. Assuming that the vanishing of the
interfacial free energy is governed by a zero-temperature FP, the Widom
exponent $\mu$ turns out to be {\em independent} of $q$ for all
sufficiently large $q$, taking the Ising value $\mu = 1$.

This is in contrast to the perturbative expansion in powers of $(q-2)$
investigated by Ludwig and Cardy \cite{Ludwig-Cardy},
Ludwig \cite{Ludwig87,Ludwig90}, and Dotsenko
{\em et al.}~\cite{Dotsenko95}. Using the RG approach for the perturbation
series around the conformal field theories representing the pure models,
these authors find the critical behaviour of the RBPM to be controlled by a
new random FP which merges with the pure FP as $q \rightarrow 2$. Critical
exponents are found to depend continuously on $q$, at least for $(q-2)$
small, and in the case of the magnetic exponent $x_H$ a calculation to
three loop order yields a prediction which is supposed to be very precise
even up to $q=3$ \cite{Dotsenko95}. Unfortunately, extending these results
beyond $q=4$ is impossible, even in principle, since this is the limiting
case in the range of minimal conformal theories around which the
perturbative calculations take place. Another interesting implication of
this line of research is that the local operators exhibit
{\em multiscaling} \cite{Ludwig90}, meaning that correlation functions of
different moments of such operators decay with powers that are, in general,
independent.

It has been suggested by Kardar \em et al. \em  \cite{Kardar95}
and one of us \cite{Cardy96} that these contrasting
theories describe very different FPs. Indeed, it can be argued that the
interface model pertains to the case of strong non-self-dual
randomness, whilst the $(q-2)$-expansion is relevant for weak self-dual
randomness. Also, even though it may turn out that the critical exponents
do not depend on $q$, the central charge $c$ evidently must, since even
when the critical behaviour is controlled by a decoupled Ising FP there is
generally not just {\em one} Ising model but several.

To resolve this controversy we have undertaken an extensive study of the $d=2$
RBPM where finite-size data obtained from transfer matrix (TM) calculations
were combined with the powerful techniques of conformal invariance.
We have extended the random cluster model TMs of Bl\"{o}te and Nightingale
\cite{Blote82} to the case of bond randomness whilst taking
into account that in the impure case such TMs do not commute and hence must
be discussed in terms of their Lyapunov (rather than the eigenvalue) spectra.
Because of the lack of self-averaging the relation between the Lyapunov
spectra and the critical exponents is inferred indirectly through a
cumulant expansion, which has the advantage of illustrating the
multiscaling properties of the correlation functions \cite{Ludwig90}
explicitly. The number of Potts states $q$ enters our TMs only as a
continuous parameter, both facilitating the comparison with analytical
results within the $(q-2)$ expansion and making the interesting regime $q>4$
readily accessible.

Although the cumulant expansion yields very appealing results in the case of
the magnetic exponent it works poorly for the thermal one. For reasons yet
not fully understood such results, when taken at face value, seem to hint
at a conformal field theory violating the bound $\nu \ge 2/d$
\cite{Chayes86,Chayes89}. On the other
hand, using phenomenological RG techniques \cite{Nightingale76}
we find results consistent both with the bound and with the
$(q-2)$-expansion \cite{Ludwig87,Dotsenko96}.

Some of our results have been reported in a Letter \cite{Jesper3}, where we
also described a mapping between the interfacial models of the RBPM for large
$q$ and the RFIM.. This mapping, which is asymptotically exact in the limit
$q \rightarrow \infty$, allowed us to establish a schematic phase diagram
for the RBPM unifying pure, non-trivial random, and percolative
behaviour. In the present paper the evidence for this phase diagram will be
collected and discussed.

The outline of this paper is as follows.
In Sect.~\ref{Sect:model} we define the model and discuss the principles of
extracting physical information from the Lyapunov spectrum of the TMs.
The proposed phase diagram is reviewed along with the translation of the
renormalisation group equations from the RFIM to the problem at hand.
Then, in Sect.~\ref{Sect:transfer}, the TM formalism of Ref.~\cite{Blote82}
is generalised to the random case. This relies on the mapping of the RBPM
to the random cluster model and on two complementary representations of the
connectivity of a row of spin in the latter model. By decomposing the TM
into sparse single-bond TMs we arrive at a highly efficient algorithm, the
implementation of which is considered in detail. The magnetic properties
can be accessed by adding a ghost site, but we also descibe an alternative
route in which the two-point correlator is related to a disorder operator
under duality. The corresponding implementation of the TMs has a seam
spanning the length of the cylinder. In the percolation limit the TMs take
on a particularly simple form, allowing us to obtain very accurate results.

Sect.~\ref{Sect:results} is dedicated to the presentation of our numerical
results. {}From the scaling of the free energy we find evidence that the
first-order phase transition is indeed softened due to the randomness. The
effective central charge is determined both at the random FP and in the
percolation limit. In the latter case we obtain excellent agreement with
our analytical result. It is shown how a cumulant expansion leads to very
accurate values of $x_H$. These depend continuously on $q$, and are in
perfect agreement with the $(q-2)$-expansion at $q=3$. For larger $q$ the
values stay far away from the Ising value, in sharp contrast to the results
of Ref.~\cite{Chen95}. For $q>8$ the expansion begins to break down, and we
give an argument why this must be so in terms of a model of coupled replicas.
The problems encountered when trying to extract $x_T$ in a similar fashion
then lead us to discuss the method of extracting physical observables from
the Lyapunov spectrum in more physical terms. We then consider the
constraints put on the multiscaling exponents by a conformal sum rule.
A reliable determination of $x_T$ is furnished by a variant of the
phenomenological RG scheme, in which the shape of the self-dual surface is
explicitly taken into account. The criticism of Ref.~\cite{Chayes86}
recently raised by P\'azm\'andi {\em et al.} \cite{Pazmandi} is shown not
to apply to the RBPM. We conclude the section with a
discussion of the higher Lyapunov spectrum and its possible relation to
the (presently unknown) conformal field theory underlying the model.

Finally, Sect.~\ref{Sect:discussion} contains a discussion of our
findings. We seek to explain the discrepancy with Ref.~\cite{Chen95}, and
we discuss other types of randomness relevant to the question whether a
first-order pahse transition is softened due to impurities. 
A list of unsettled questions relevant for future research is also given.

\section{The model and its phase diagram}
\label{Sect:model}

\subsection{The random bond Potts model}

The $q$-state Potts model \cite{Potts52} is defined by the reduced
Hamiltonian
\begin{equation}
  {\cal H} = - \sum_{\langle ij \rangle} K_{ij} \delta_{\sigma_i \sigma_j},
  \label{H}
\end{equation}
where the spins, defined on the vertices of the square lattice,
can take the values $\sigma_i = 1,2,\ldots,q$, and the summation is over all
nearest neighbour bonds in the lattice. We shall specialise to the
ferromagnetic case, where the reduced couplings $K_{ij} \ge 0$ measure the
strength of the aligning tendency of nearest neighbour spins.

Although the free energy of the pure model ($K_{ij} \equiv K$) is not known
in closed form for general $q$, a wide range of exact results is nevertheless
available \cite{Wu82}. In particular it is well-known that the model
exhibits a second-order phase transition for $q \le 4$ and a first-order
one for $q > 4$ \cite{Baxter73}.

However, in this paper we are mainly concerned with
the random bond Potts model (RBPM) for which much less is known. Here the
couplings $K_{ij}$ are quenched random variables, typically drawn from the
symmetric binary distribution
\begin{equation}
  P(K) = \frac{1}{2}[ \delta(K-K_1) + \delta(K-K_2) ],
  \label{binary}
\end{equation}
where the ratio between strong and weak bonds $R = K_2 / K_1$ measures the
strength of the randomness. For the special choice
\begin{equation}
  \left( {\rm e}^{K_1} - 1 \right) \left( {\rm e}^{K_2} - 1 \right) = q
  \label{selfdual}
\end{equation}
the model is on average self-dual, as discussed in more detail in
Sect.~\ref{Sect:magnetic} below. Assuming that the phase transition is
unique the model is therefore at its critical point \cite{Kinzel81}.

Other self-dual distributions of the random bonds than that of
Eq.~(\ref{binary}) have also been investigated in order to check our
results. In particular, we have found the trinary distribution introduced in
Sect.~\ref{Sect:central-charge} useful, since it gives us a clearer idea
about the length scale associated with the random impurities.

\subsection{Lyapunov spectrum of the transfer matrix}

The construction of the transfer matrices (TMs) for the RBPM is described in
detail in Sect.~\ref{Sect:transfer}. It is well-known that in the pure case
($R=1$) the operator content of the conformal field theory (CFT) underlying the
model is related to the eigenvalue spectrum
$\{\lambda_i(L)\}$, $i=0,1,2,\ldots$, 
of the TM for a strip of width $L$ through \cite{Cardy83}
\begin{equation}
  f_i(L) - f_0(L) = \frac{2\pi x_i}{L^2} + \cdots,
  \label{FSS-x}
\end{equation}
where $f_i(L) = -\frac{1}{L}\ln \lambda_i(L)$ are the generalised free energies
per site (in units of $k_{\rm B} T$) and $x_i$ the scaling dimensions of
the corresponding operators. 
Similarly the central charge $c$, measuring the number of bosonic degrees of
freedom of the CFT, is related to the finite-size corrections to the
customary free energy through \cite{Cardy86}
\begin{equation}
  f_0(L) = f_0(\infty) - \frac{\pi c}{6L^2} + \cdots.
  \label{FSS-c}
\end{equation}

In the random case the TMs are no longer constant but depend on the
particular realisation of the random bonds within each row of
strip. Accordingly the concept of eigenvalues generalises to that of
Lyapunov exponents.
Starting with some suitable initial vector of unit norm
$|{\rm v}_0 \rangle$, the leading Lyapunov exponent can be found by
the Furstenberg method \cite{Furstenberg}
\begin{equation}
  \Lambda_0(L) = \lim_{m \rightarrow \infty} \frac{1}{m}
                 \ln \left \| \left( \prod_{j=1}^m {\cal T}_j
                 \right) |{\rm v}_0 \rangle \right \|,
  \label{Furstenberg}
\end{equation}
where ${\cal T}_j$ is the TM acting between rows $j-1$ and $j$.
The average free energy per site is given as before by
$f_0(L) = - \frac{1}{L} \Lambda_0(L)$.
Higher exponents are found by iterating a set of $n$ vectors
$\{ |{\rm v}_k \rangle \}_{k=0}^{n-1}$, where a given
$|{\rm v}_k \rangle$ is orthogonalised to the set
$\{ |{\rm v}_l \rangle \}_{l=0}^{k-1}$ after each
multiplication by ${\cal T}_j$ \cite{Benettin}. Surprisingly, this method
works even for a non-hermitian TM, and it is numerically shown to be
independent of the choice of the initial vectors.

When some symmetry (e.g., spin reversal or duality)
is manifest in ${\cal T}_j$ the orthogonalisation can be circumvented by
iterating vectors which belong to definite irreducible components of
that symmetry, but the $S_q$ permutational symmetry inherent in the
Potts model has been lost through the mapping to the random cluster
model which forms the backbone of our TMs; see Sect.~\ref{Sect:random-cluster}.

As to the extraction of physical information from the spectra,
Eq.~(\ref{FSS-c}) is supposed to retain its validity provided that $c$ is
replaced by the effective central charge $c'$, that in the standard replica
formalism is the derivative of $c(n)$ with respect to the number of
replicas at $n=0$ \cite{Ludwig-Cardy}. The question to which extent
Eq.~(\ref{FSS-x}) also remains valid is by no means trivial and we shall
dedicate a fair part of the subsequent discussion to it.

\subsection{Phase diagram}
\label{Sect:phase-diagram}

In the limit $q \rightarrow \infty$ the behaviour of the pure model is
readily understood \cite{Jesper3}. At the self-dual point the partition
function is dominated by two contributions, namely those corresponding to
the $q$ completely ordered states and the completely disordered
state respectively. All other 
configurations are down by powers of $1/\sqrt{q}$ and have recently been
enumerated to 10th order in this small parameter \cite{Bhattacharya}.
The dominating states have identical free energy but different internal
energy densities of $K$ and $0$ for the ordered and the disordered phase
respectively, so the transition is, as expected, first order.

The bond randomness is then included through the parametrisation
${\rm e}^{K_{ij}}-1 = q^{\frac{1}{2}+w_{ij}}$, where $w_{ij} = \pm w$ and $w>0$
measures the strength of the randomness. It can now be shown \cite{Jesper3}
that as $q \rightarrow \infty$ the model for an interface between these two
phases of the RBPM is exactly the same as that of an interface between the
spin-up and spin-down phases of the RFIM
\begin{equation}
  {\cal H}^{\rm RFIM} =
  - J \sum_{\langle ij \rangle} s_i s_j 
  - \sum_i h_i^{\rm RF} s_i - h \sum_i s_i
\end{equation}
with $h_i^{\rm RF} = \pm h^{\rm RF}$, provided that one
translates quantities between the two models using the ``dictionary''
\begin{eqnarray}
  h^{\rm RF} & \leftrightarrow & \frac{1}{2} w \ln q \nonumber \\
  J          & \leftrightarrow & \frac{1}{8} \ln q \\
  h          & \leftrightarrow & \frac{1}{4} t \ln q \nonumber.
\end{eqnarray}
Here $t \equiv (T-T_{\rm c})/T_{\rm c}$ is the reduced temperature.

The infinitesimal RG equations can now be inferred from the similar
results for the RFIM \cite{Bray-Moore}. Near $d=2$ they read
\begin{eqnarray}
  dw/dl            &=& -(d/2-1)w + Aw^3 + \cdots \label{rg1} \\
  d(\ln q)^{-1}/dl &=& -(\ln q)^{-1}\big((d-1)-Aw^2+\cdots\big)\label{rg2}\\
  dt/dl            &=& t(1+Aw^2+\cdots), \label{rg3}
\end{eqnarray}
where $A>0$ is a non-universal constant.
The RG flows for $d>2$ and the proposed phase diagram are shown in
Fig.~\ref{Fig:phase-diagram} (see Ref.~\cite{Jesper3} for details). The
shaded region of non-vanishing latent heat is bounded by the line $Rq_2$
of tricritical points. This line, controlled by the fixed point $R$ at
infinite $q$, merges with the abscissa as $d \rightarrow 2$. At 
$q = \infty$ the interfacial mapping is exact so that the flows along
$RP_1$ must extend all the way to $w = \infty$, and since $w^{-1}$
is known to be a relevant scaling variable at the percolation limit
\cite{percolation}, Ref.~\cite{Jesper3} concluded that $Rq_2$ must be
separated from the percolative behaviour along $P_1P_2$ by another line of
stable FPs emerging from $P_1$.
It was then conjectured that this connects on to the line of random FPs
found in the $(q-q_1)$-expansion \cite{Ludwig-Cardy}. In this paper we
shall present the evidence for this conjecture for $d=2$, when $q_1 = 2$
and $q_2 = 4$.

\section{The transfer matrices}
\label{Sect:transfer}

In spite of the large amount of high-precision results obtained by
combining transfer matrix (TM) techniques with finite-size scaling for
almost any conceivable type of pure statistical mechanics system (see,
e.g., Ref.~\cite{FSS} for a review) the use of TMs in the
study of disordered systems seems to have attracted rather little
interest as compared with the complementary approach of Monte Carlo
simulations. 

A straightforward way of setting up the TMs for the
$q$-state Potts model is to use the traditional spin basis where the state
of a row of $L$ spins is labelled by the $q^L$ basis states
$\lbrace \sigma_1, \sigma_2, \ldots, \sigma_L \rbrace$,
$\sigma_i = 1, \ldots, q$. Whilst
this approach is highly efficient for $q=2,3$ it has two major shortcomings
in the general case. First, the dimension of the matrices becomes
forbiddingly for large $q$, in particular making unaccessible the
regime of $q>4$ which is our main concern. Second, the restriction to
integer values of $q$ is unnecessary and in fact makes it difficult to
compare numerical results with analytical calculations in the
$(q-2)$-expansion \cite{Ludwig87,Ludwig90,Dotsenko95}.

Both these shortcomings can be remedied by writing the TMs in the 
connectivity basis introduced by Bl\"{o}te and Nightingale
\cite{Blote82}. In this representation the dimension of the TMs is
independent of $q$ which enters only as a {\em continuous} parameter.
In fact, the number
of basis states is asymptotically $\sim 4^L$ (or $\sim 5^L$ upon
imposition of a magnetic field) with a rather small coefficient of
proportionality, in practice making this basis the preferred choice
for all but the Ising model ($q=2$).

We have generalised these TMs to include quenched bond
randomness, and also devised an alternative method of accessing the
magnetic properties through the introduction of a seam along the
strip. Furthermore, in the percolation limit the TMs are found to
simplify in a manner that makes calculations for rather large strip
widths feasible. For convenience these results are presented along
with a review of the relevant parts of Ref.~\cite{Blote82} thus making
our description of the Potts model TMs self-contained.

\subsection{Mapping to the random cluster model}
\label{Sect:random-cluster}

Introducing an imaginary `ghost site' with fixed spin $\sigma_0 = 1$
the partition function for the Potts model can be written as
\begin{equation}
  Z = \sum_{\lbrace \sigma \rbrace}
      \left( \prod_{\langle ij \rangle}
             \exp(K_{ij} \delta_{\sigma_i \sigma_j}) \right)
      \left( \prod_{\langle i0 \rangle}
             \exp(H_i \delta_{\sigma_i \sigma_0}) \right),
\end{equation}
where $\prod_{\langle ij \rangle}$ is the usual product over pairs of nearest
neighbour sites and each site $i$ has been connected to the ghost
site $0$ with a similar notation.
The reduced magnetic field $H_i$, here taken to be site dependent, now
enters at the same footing as the reduced exchange couplings $K_{ij}$.
It should be pointed out, however, that a random coupling to the ghost site
is not a true random field, since the latter would try to force different
sites into {\em different} Potts states and not just into the particular
state of the ghost site with a site-dependent probability. To avoid any
confusion we shall therefore specialise to the case of a homogeneous field
$H_i \equiv H$.

The site variables can now be traded
for bond variables through the mapping to the random cluster model
introduced by Kasteleyn and Fortuin \cite{Kasteleyn}. In terms of the
variables $u_{ij} = {\rm e}^{K_{ij}} - 1$ and $v = {\rm e}^{H} - 1$
we arrive at
\begin{equation}
  Z = q^N \sum_{G \subseteq {\cal L}} \sum_{G_0 \subseteq {\cal L}_0}
      \left( \prod_{\langle ij \rangle \in G} \frac{u_{ij}}{q} \right)
      \left( \prod_{\langle i0 \rangle \in G_0} \frac{v}{q} \right)
      q^{l(G \cup G_0)},
  \label{cluster-model}
\end{equation}
where ${\cal L}$ denotes the set of all nearest neighbour bonds,
${\cal L}_0$ the bonds from each of the $N$ Potts spin to the ghost
site, and $l(G \cup G_0)$ is the number of independent loops on the
combined graph $G \cup G_0$.

The usual construction of the transfer matrix ${\cal T}$ for a strip of
width $L$ seems to be obstructed by the non-local factor $l(G \cup G_0)$, but
this can be taken into account by choosing a basis containing
information about which sites of a given row are interconnected through
the part of the lattice below that row (including connections via the
ghost site). This leads us to the concept of connectivity states,
which we consider next.

\subsection{The connectivity states}

In order to determine the number of loop closures induced by appending a
new row of $L$ sites along with the corresponding $L$ connections to the
ghost site to the top of $G \cup G_0$, we need information
about how the sites in the top row of $G \cup G_0$ were previously
interconnected. This information is comprised in the {\em connectivity
state} $(i_1 i_2 \ldots i_L)$, where $i_t = 0$ if site $t$ is
connected to $0$ within the combined graph $G \cup G_0$ and, otherwise,
$i_r = i_s$ is a (non-unique) positive integer if and only if sites 
$r$ and $s$ are connected within $G$.

Whilst this `index representation' is useful for determining whether
a newly appended bond does or does not close a loop, and thus will
allow us to explicitly construct the single-bond TMs in the next
subsection, a one-to-one mapping to the set of consecutive integers
$\lbrace 1, 2, \ldots \rbrace$ is clearly needed to define a `number
representation' which will enable us to label the entries of the TM and
thus to perform actual computations. These representations and the
mapping were supplied by Ref.~\cite{Blote82} as were the determination of
the number of connectivity states ($d_L$ with and $c_L$ without a
magnetic field). We shall review the necessary details and also give
details on the construction of the inverse of the mapping just mentioned. 

Consider first the case of $H = 0$ where all ghost bonds carry zero
weight ($v=0$). The connectivity states then have all $i_t > 0$ and can
be recursively ordered by noting that the index representation is {\em
well-nested}, i.e., for $r < s < t < u$
\begin{equation}
  (i_r = i_t) \land (i_s = i_u) \Rightarrow i_s = i_t.
\end{equation} 
It follows that if we define the {\em cut function}
$\rho(i_1 i_2 \ldots i_L)$ to be the smallest $t>1$ such that $i_1 = i_t$,
if such a $t$ exists, and $L+1$ otherwise, the left
$(i_2 i_3 \ldots i_{\rho-1})$ and right $(i_{\rho} i_{\rho+1} \ldots i_L)$
parts of the index representation are both well-nested. A complete
ordering of the well-nested sequences is now induced by applying the
cut function first to the whole sequence, then recursively to its
right and finally to its left part.

More precisely, the mapping from the index to the number representation is
effected by
\begin{equation}
  \sigma(i_1 i_2 \ldots i_L) = \left \{ \begin{array}{l}
    1 \mbox{ if } L \le 1 \\
    \begin{array}{r}
      c_{L,k-1} + [\sigma(i_k \ldots i_L) - 1] c_{k-2} \\
      + \sigma(i_2 \ldots i_{k-1}) \mbox{ otherwise,}
    \end{array}
    \end{array} \right.
  \label{sigma}
\end{equation}
where $k = \rho(i_1 \ldots i_L)$ and
$c_{n,l} = \sum_{i=2}^l c_{i-2} c_{n-i+1}$ 
with
\begin{equation}
  c_n \equiv c_{n,n+1} = \frac{(2n)!}{n!(n+1)!}
\end{equation}
giving the number of well-nested $n$-point connectivities
\cite{Blote82,Blote89}. Explicit values are shown in Table \ref{Tab:cn}.

To consider the general case of $v \neq 0$ we remark that the
subsequence of non-zero indices is still well-nested. A complete
ordering of an index representation $(i_1 i_2 \ldots i_L)$ with precisely $s$
zero indices is then induced by first ordering according to the value
of $s$, then lexicographically ordering the zeros, and finally using
the ordering of the well-nested subsequence
$(i_{p_1} i_{p_2} \ldots i_{p_{L-s}})$ given by Eq.~(\ref{sigma}).
The lexicographic ordering is carried out by
\begin{equation}
  \psi(i_1 i_2 \ldots i_L) = \left \{ \begin{array}{l}
    1 \mbox{ if } L=1 \mbox{ or } s=L \\
    \psi(i_2 i_3 \ldots i_L) \mbox{ if } i_1 \neq 0 \\
    {L-1 \choose s} + \psi(i_2 i_3 \ldots i_L) \mbox{ if } i_1 = 0,
  \end{array} \right.
\end{equation}
and the mapping to the number representation is finally
\begin{eqnarray}
  \tau(i_1 i_2 \ldots i_L) = d_{L,s-1}
    &+& [\psi (i_1 i_2 \ldots i_L) - 1] c_{L-s} \nonumber \\
    &+& \sigma (i_{p_1} i_{p_2} \ldots i_{p_{L-s}}),
  \label{tau}
\end{eqnarray}
where $d_{n,l} = \sum_{i=0}^l {n \choose i} c_{n-i}$ with
\begin{equation}
  d_n \equiv d_{n,n} = \sum_{i=0}^n {n \choose i} c_{n-i}
\end{equation}
giving the number of general $n$-point connectivities. Again, explicit
values are presented in Table \ref{Tab:cn}.

To construct the inverse mapping, i.e., the one taking us from the
number to the index representation, we solve $\tau = \tau(i_1 i_2 \ldots i_L)$
for the indices $(i_1 i_2 \ldots i_L)$ by performing the following steps.
First, the number of zero indices is found as
$s = \max \{s | d_{L,s-1} < \tau \}$. Second, perform a slightly modified
integer division by writing $\tau - d_{L,s-1} = Q c_{L-s} + R$,
where the remainder $R$ is restricted to take its values in the interval
$[1,c_{L-s}]$. {}From Eq.~(\ref{tau}) we infer that $\psi = Q+1$ and
$\sigma = R$. 
Third, the position of the first (leftmost) zero index is given by
$i_{L-k} = 0$, where $k = \max \{k | {k \choose s} < \psi \}$.
This procedure of finding the zero indices is then iterated with
$\psi \rightarrow \psi^{(1)} \equiv \psi - {k \choose s}$ until
$\psi^{(s')} = 1$, and the remaining $s - s'$ zeros are filled in from
the right: $i_{s'+1} = \cdots = i_{s-1} = i_s = 0$.

It remains to deduce the subsequence of non-zero indices by inverting
$\sigma = \sigma(i_{p_1} i_{p_2} \ldots i_{p_l})$ with $l = L-s$. After
initialising $i_{p_1} = p_1$ we proceed by recursion as follows.
First, choose $k = \min \{k | c_{l,k-1} + c_{k-2} c_{l-k+1} \ge \sigma \}$.
If $k \le l$ we have then found a connection: $i_{p_1} = i_{p_k}$.
This procedure of finding the connections is now iterated on the
left $(i_{p_2},\ldots,i_{p_{k-1}})$ and the right
$(i_{p_k},\ldots,i_{p_l})$ parts of the remaining sequence.
If $k \ge 2$ the assignment $i_{p_2} = p_2$ is performed.
By (modified) integer division we then write
$\sigma - c_{l,k-1} = Q c_{k-2} + R$ with $R \in [1,c_{l-k+1}]$, and pass
over the left part of the sequence with
$\sigma \rightarrow \sigma^{(1)}_{\rm left} \equiv R$ and
$\l \rightarrow \l^{(1)}_{\rm left} \equiv k-2$, and the right part with
$\sigma \rightarrow \sigma^{(1)}_{\rm right} \equiv Q+1$ and
$\l \rightarrow \l^{(1)}_{\rm right} \equiv l-k+1$.
The recursion stops when for any sequence $l^{(m)} \le 2$. If then
$l^{(m)} = 2$ and the sequence is $(i_{p_a},i_{p_{a+1}})$ we perform the
assignment $i_{p_{a+1}} = i_{p_a}$ if $\sigma^{(m)} = 1$ and
$i_{p_{a+1}} = p_{a+1}$ if $\sigma^{(m)} = 2$.

Any way of constructing the index representation
$(i_1 i_2 \ldots i_L)$ will of course reflect the above-mentioned 
arbitrariness as to the actual values of the non-zero indices,
but the particular procedure just outlined is easily seen to ensure
that all indices are $\le L$. This invariant is useful since then any given
site $t$ can be {\em disconnected} from the rest by assigning $i_t = L+1$.

\subsection{The single-bond transfer matrices}
\label{Sect:single-bond}

The amount of computer time necessary for building up a long strip by
repeated application of the transfer matrix ${\cal T}$ can be enormously
reduced by decomposing the latter as a product of sparse matrices, each
corresponding to the addition of a single bond to ${\cal L}$.

Specifically we write
${\cal T} = {\cal T}^0 {\cal T}^{\rm h} {\cal T}^{\rm v}$,
where
${\cal T}^{\rm v} = {\cal T}^{\rm v}_L \cdots
 {\cal T}^{\rm v}_2 {\cal T}^{\rm v}_1$
is connecting each of the $L$ spin sites in the uppermost row of the strip
to a new spin site situated {\em vertically} above it, and
${\cal T}^{\rm h} = {\cal T}^{\rm h}_{L,1} \cdots
 {\cal T}^{\rm h}_{2,3} {\cal T}^{\rm h}_{1,2}$
is finishing the new row of ${\cal L}$ by appending {\em horizontal} bonds
between each of the nearest-neighbour dangling ends created by
${\cal T}^{\rm v}$. The matrix
${\cal T}^{\rm h}_{L,1}$ imposes periodic boundary conditions by
interconnecting the newly added spins at sites $L$ and 1. Finally
${\cal T}^0 = {\cal T}^0_L \cdots {\cal T}^0_2 {\cal T}^0_1$
furnishes the bonds of ${\cal L}_0$ from each of the new spin sites
to the ghost site.
Each of these single-bond TMs is implicitly understood to depend on the
particular realisation of the bond and, in the case of ${\cal T}^0_i$, the
field randomness pertaining to the bond in question.

Upon addition of one single bond the summation over graphs 
in Eq.~(\ref{cluster-model}) is augmented by a sum over the two possible
states of this new degree of freedom, viz.~the bond added to ${\cal L}$
(${\cal L}_0$) can be either {\em present} or {\em absent} in $G$
($G_0$). Correspondingly each column of the TM has at most two distinct
non-zero entries.

Consider first adding a vertical bond by action of 
${\cal T}^{\rm v}_l$, $l \in \{1,\ldots,L\}$. If the bond is `present'
any given connectivity state $(i_1 i_2 \ldots i_L)$ of the $L$ uppermost
spin sites will be left unchanged. In case of an `absent' bond site $l$
will be disconnected, and the number representation of the new connectivity
state can be found by assigning $i_l = L+1$ and using
Eq.~(\ref{tau}). Interpreting the factor of $q^N$ in Eq.~(\ref{cluster-model})
as an extra factor of $q$ going with each vertical bond we see that the
non-zero entries in ${\cal T}^{\rm v}_l$ corresponding to a column
with a given connectivity number are a diagonal
contribution of $u_{ij}$ and a possibly off-diagonal contribution of
$q$. In particular the vertical bonds do not induce any loop closures.

Similarly the TM of a horizontal bond ${\cal T}^{\rm h}_{l,l+1}$ has
a diagonal entry of $1$ for each column, corresponding to the bond being
absent. The other non-zero entry corresponds to a present bond, and its
value depends on whether a loop is being closed or not. Given the
connectivity state $(i_1 \ldots i_l i_{l+1} \ldots i_L)$ of some column in
the TM this is determined by comparing $i_l$ and $i_{l+1}$: if they are
equal we get an additional diagonal contribution of $u_{l,l+1}$
corresponding to a loop closure, whereas if they are different there is an
off-diagonal entry with value $u_{l,l+1} / q$. In the latter case the
connectivity number is found by assigning the value $\min \{i_l,i_{l+1} \}$
to all indices that were formerly equal to either $i_l$ or $i_{l+1}$ and
applying Eq.~(\ref{tau}). (The reason why we copy the {\em minimum} index is to
ensure the proper handling of spins connected to the ghost site.)

Finally, the TM of a ghost bond ${\cal T}^{\rm h}_l$ has the same
form as in the case of a horizontal bond if we make the substitutions
$u_{l,l+1} \rightarrow v_l$ and $i_{l+1} \rightarrow 0$.

\subsection{Magnetic properties}
\label{Sect:magnetic}

It is well known, at least in the case of a pure system, that physically
interesting quantities like the central charge $c$ as well as the thermal
($x_T$) and the magnetic ($x_H$) scaling dimensions can be extracted from
the transfer matrix spectrum. Consider for the moment the case of vanishing
magnetic field, $H=0$. Since connections to the ghost site are then
generated with zero weight ($v=0$) such connections can only be present in
any row if they were already there in the preceding row. In particular,
noting that in the numbering of connectivities induced by Eq.~(\ref{tau})
the non-ghost connectivities precede the others, we see that the TM assumes
the following block form \cite{Blote82}
\begin{equation}
  {\cal T} = \left[ \begin{array}{cc}
                {\cal T}^{11} & {\cal T}^{12} \\
                0             & {\cal T}^{22}
             \end{array} \right],
\end{equation}
where superscript 2 (1) refers to the (non-)ghost connectivities.

The largest and the next-largest eigenvalues of ${\cal T}$ turn out to be
the largest eigenvalue of block ${\cal T}^{11}$ and ${\cal T}^{22}$
respectively, and from the corresponding (reduced) free energies per site
$f_0^{ii}(L) = - \frac{1}{L} \lambda_0^{ii}$ ($i=1,2$) for a strip of
width $L$ the magnetic scaling dimension can be found from the CFT formula
\cite{Cardy83}
\begin{equation}
  f_0^{22}(L) - f_0^{11}(L) = \frac{2 \pi x_H}{L^2} + \cdots.
  \label{xh}
\end{equation}
Physically this relation to $x_H$ can be understood by noting that by
acting repeatedly with ${\cal T}^{22}$ on some initial (row) state
$|{\rm v}_0 \rangle \neq 0$ one measures the decay of clusters
extending back
to row 0. This must have the same spatial dependence as the spin-spin
correlation function and hence be related to $x_H$ \cite{Blote82}.
Analogously ${\cal T}^{11}$ measures the decay of two-point correlations
between {\em pairs} of spins being interconnected within the random cluster
model. This is nothing but the energy-energy correlation in the strip
geometry, and accordingly we expect that
\begin{equation}
  f_1^{11}(L) - f_0^{11}(L) = \frac{2 \pi x_T}{L^2} + \cdots.
  \label{xt}
\end{equation}

We have checked the results for $x_H$ by constructing a realisation of the
TM in the presence of a {\em seam} spanning the length of the cylinder. Our
algorithm also merits attention on its own right since it improves the
asymptotic number of basis states necessary for finding $f_0^{22}(L)$ from
$d_L - c_L \sim 5^L$ (the dimension of ${\cal T}^{22}$) to
$L c_L \sim L 4^L$. In practice, however, with the strip widths $L$
accessible using present-day computers the two algorithms perform more or
less equally fast (see Table~\ref{Tab:cn} for a comparison).

The well-known duality relation for the Ising model partition function
without a magnetic field is easily extended to the case of the Potts model
on a cylinder.
For $v = 0$ the partition function of the random cluster model,
Eq.~(\ref{cluster-model}), can be rewritten as
\begin{equation}
  Z = \sum_{G \subseteq {\cal L}}
      \left( \prod_{\langle ij \rangle \in G} u_{ij} \right) q^{C(G)},
\end{equation}
where $C(G)$ is the number of independent clusters on $G$.
We first stipulate the duality between two very special graphs. Namely,
the full graph $G = {\cal L}$ with partition function
$Z_{\rm full}(\{u_{ij}\}) = q \prod_{\langle ij \rangle} u_{ij}$
is taken to be dual to the empty graph $G^{\ast} = \emptyset$ with
$Z^{\ast}_{\rm empty}(\{u_{ij}^{\ast}\}) = q^{N^{\ast}}$,
where the number of dual sites $N^{\ast}$ is fixed by the Euler relation.

Establishing the duality then amounts to ascertaining that all other graphs
have the same weight relative to this reference state as is the case in the
dual model. In the terminology introduced above,
duality states that a graph configuration $G$ on the original lattice
${\cal L}$ is dual to a configuration $G^{\ast}$ on the dual lattice
${\cal L}^{\ast}$ in which every bond of strength $u_{ij}$ being `present'
in $G$ corresponds to the dual bond of strength $u_{ij}^{\ast}$ being
`absent' from $G^{\ast}$.

In particular, removing one bond from the full graph (relative weight:
$1 / u_{ij}$) must correspond to adding the corresponding dual
bond to the empty dual graph (relative weight: $u^{\ast}_{ij} / q$),
meaning that the bond strengths and their duals must obey the relation
\begin{equation}
  u_{ij} u^{\ast}_{ij} = q.
\end{equation}
When removing further bonds from $G$ it may happen that a new cluster is
separated from the rest of the graph, yielding an additional factor of
$q$. But such a cluster formation corresponds precisely to a loop closure
on the dual lattice, also giving an extra factor of $q$! Since all graph
configurations $G$ can be constructed by succesive removals of bonds from
the full reference state we have thus proven the fundamental duality
relation \cite{Wu97a}
\begin{equation}
  Z(\lbrace u_{ij} \rbrace) = qC Z^{\ast}(\lbrace u^{\ast}_{ij} \rbrace),
\end{equation}
where $C = q^{-N^{\ast}} \prod_{\langle ij \rangle} u_{ij}$ is some constant.

A similar duality relation can be established for the spin-spin correlation
function. As usual we define the local order parameter as \cite{Cardy-book}
\begin{equation}
\label{orderparameter}
  M_a(r) = \left( \delta_{\sigma(r),a} - \frac{1}{q} \right),
  \mbox{ $a=1,\ldots,q$}.
\end{equation}
In the high temperature phase all components of the order parameter vanish,
whilst in the ordered (low temperature) phase the $Z_q$ symmetry is
spontaneously broken and one of the components, say $a=1$, has a positive
expectation value.
A simple calculation now shows that the correlation function
$G_{aa}(r_1,r_2) = \langle M_a(r_1) M_a(r_2) \rangle$ is 
proportional to the probability that the points $r_1$ and $r_2$ belong to the
same cluster.

In a cylindrical geometry the graphs with $r_1$ and $r_2$, taken to be at
opposite ends of the cylinder, connected correspond to dual graphs where
clusters are forbidden to wrap around the cylinder. This is equivalent
to computing the dual partition function with twisted boundary conditions
$\sigma \rightarrow (\sigma+1)$ mod $q$ across a {\em seam} running from
$r_1$ to $r_2$. By permuting the Potts spin states the shape of this seam
can be deformed at will as long as it connects $r_1$ and $r_2$.
Duality thus maps the correlation function onto a disorder operator
\begin{equation}
  \langle M_a(r_1) M_a(r_2) \rangle = 
  \left \langle \prod_{\rm seam} \exp(-K^{\ast} \delta_{\sigma_i
    \sigma_j}) \right \rangle_{Z^{\ast}},
  \label{disorder-op}
\end{equation} 
where $Z^{\ast} = Z^{\ast}(\{K^{\ast}\})$ is the dual partition function
with periodic boundary conditions.

The construction of the TM in the presence of a seam is facilitated by the
following observation: If no cluster is allowed to wrap the cylinder,
each graph contributing to the partition function can be associated with
a function $s(j)$ of the row number $j$, such that
$s(j)=k \in \{1,\ldots,L\}$ means that in row $j$ no horizontal bond
connecting sites $k$ and $k+1$ (mod $L$) is present. For obvious reasons we
shall refer to $s$ as the {\em virtual seam}. We can then write the TM in a
basis which is the direct product of the $L$ possible values of the virtual
seam and the costumary $c_L$ non-ghost connectivities. The virtual seam is
initialised by assigning to it a definite value in row 0, viz.~$s(0) = L$
for all graph configurations of that row.

The single-bond TM of a vertical bond is diagonal in $s$, but a present
horizontal bond {\em not} inducing a loop closure may alter the value of
the virtual seam. Let us recall from Sect.~\ref{Sect:single-bond} that to
find the connectivity state $(i_1 \ldots i_l i_{l+1} \ldots i_L)$
giving the row label of $T^{\rm h}_{l,l+1}$ that corresponds to the
off-diagonal entry with value $u_{l,l+1}/q$ we would join the two distinct
clusters formerly labelled by either $i_l$ or $i_{l+1}$. But such a merger
would ruin the invariant stated above, unless we move the virtual seam at
the same time. On the other hand, if $i_l = i_{l+1}$ and $s(j) = l$ we must
explicitly prevent a cluster from wrapping the cylinder by leaving out that
extra diagonal contribution which would otherwise by implied by the
condition $i_l = i_{l+1}$. In this case the virtual seam is not moved.

To conclude this section we remark that in the case of a {\em planar}
geometry any $n$-point Potts correlation function can be mapped to a
generalised surface tension by duality \cite{Wu97a,Jesper2,Wu97b}.

\subsection{The percolation limit}
\label{Sect:T-percolation}

In the random bond Potts model the couplings $u_{ij} \ge 0$ are quenched random
variables, and the critical point can be accessed by drawing them from the
symmetric binary distribution
$P(u) = \frac{1}{2}[\delta (u-u_1) + \delta (u-u_2)]$,
where $u_1 u_2 = q$. For details, see Sect.~\ref{Sect:results}.
Bond percolation can be studied in the limit $u_1 \rightarrow 0$, $u_2
\rightarrow \infty$ of infinitely weak and strong bonds respectively. In this
limit considerable simplifications occur in the TM, rendering computations
with rather large strip widths feasible.

In the percolation limit all single-bond TM have only one non-zero entry
per column. Recall from Sect.~\ref{Sect:single-bond} that in the general
case there are two such entries of which one is diagonal and the other is
`non-trivial'. In the case of the strong vertical bonds and the weak
horizontal bonds only the diagonal entries survive, so that the matrices
${\cal T}^{\rm v}_{\rm strong} = u_2 {\bf 1}$ and
${\cal T}^{\rm h}_{\rm weak} = {\bf 1}$
both become trivial. On the other hand, a weak vertical bond corresponds to
a TM having one non-trivial entry of $q$ per column, whilst a strong
horizontal bond is represented by a TM that is $u_2$ times a non-trivial
matrix with entries of $1$'s and $1/q$'s.

The factors of $u_2$ multiplying both
${\cal T}^{\rm v}_{\rm strong}$ and
${\cal T}^{\rm h}_{\rm strong}$
are innocuous albeit infinite, since of the $2L$ single-bond matrices
constituting the
entire ${\cal T}$ there will on average be $L$ strong ones, hence $L$
factors of $u_2$. On the level of the specific free energy this
amounts to an infinite additive constant
\begin{equation}
  f_0^{11}(L) = - \ln u_2 + \tilde{f}_0^{11}(L)
  \label{inf-const}
\end{equation}
independent of the strip width $L$. In particular, the central charge $c$
can be extracted from the {\em finite} quantity $\tilde{f}_0^{11}(L)$.

As we shall see in Sect.~\ref{Sect:results} this quantity can be found by
measuring the asymptotic growth of the norm of
$\left( \prod_{j=1}^m {\cal T}^{11}_j \right) |{\rm v}_0 \rangle$,
where $|{\rm v}_0 \rangle$ is some largely arbitrary initial
vector. In the percolation limit the TM turn out to be so sparse that after
a very few iterations the resulting vector has only one non-zero
component. Computationally this 
means that it is sufficient to store the row index of that non-zero
component as well as its norm. Both time and memory requirements are thus
enormously reduced, allowing us to access larger system sizes.

The disadvantage of this projective quality of the percolation point TM is
that neither the thermal nor the magnetic scaling dimensions can be
found from the Lyapunov spectrum. 
In the case of $x_H$ an initial vector in the ${\cal T}^{22}$ sector
will rapidly decay to zero, thus invalidating the procedure for finding
$\tilde{f}_0^{22}(L)$, and the alternative of using a seam is obstructed by
the fact that disallowing the entry in the horizontal bond TM that
corresponds to a cluster wrapping the cylinder is imcompatible with the
argument of pulling out an overall factor of $u_2$ from the TM.

\section{Results}
\label{Sect:results}

\subsection{Softening of the transition}

Before attempting to determine the universality classes of the RBPM it is
essential to make sure that quenched bond randomness indeed renders the
phase transitions second order.
For $q>4$ the pure system has a first-order transition for which the free
energy per site is expected to scale like \cite{Blote82}
\begin{equation}
  f_0(L) = f_0(\infty) + a L^{-d} \exp (-L/\xi),
\end{equation}
where $\xi$ is the bulk correlation length and $a$ is an amplitude depending
on $q$.
In Fig.~\ref{Fig:12} we show plots of the function
\begin{eqnarray}
  \lambda(L) & \equiv & \ln[ f_0(L) - f_0(\infty)] + d \ln L \\
             & \sim   & \mbox{\rm const} - L/\xi \nonumber
\end{eqnarray}
for various values of $q$ and the randomness strength $R$. These plots are
rather sensitive to the value of $f_0(\infty)$, but although this is only
known exactly for the pure model \cite{Baxter73} it can nevertheless be
determined with sufficient accuracy from the parabolic fits described in
Sect.~\ref{Sect:central-charge} below.

For $q=8$ the finite correlation length of the pure system
($\xi \sim 70$) is seen to be rendered effectively
infinite ($\xi \sim 10^3$) upon
imposition of the randomness, whilst the transition of the Ising model
($q=2$) simply stays second order.
Despite the simplicity of these plots we also find a fair
agreement with the recently found analytical values of $\xi$ for the pure
systems \cite{Buffenoir}; near $q=4$ these assume the simple form
\begin{equation}
  \xi \simeq \frac{\sqrt{2}}{8} \exp \left( \frac{\pi^2}{\sqrt{q-4}} \right).
  \label{xi-pure}
\end{equation}

Another criterion for distinguishing between first and second-order phase
transitions is the values of the (effective) exponents $x_H$ and $x_T$ as
found from Eq.~(\ref{xh}) and (\ref{xt}) respectively. Generally speaking,
for pure systems with $q>4$ these equations give rise to rather poor fits
which however have extrapolated values of the effective exponents that are
in the vicinity of, and slightly below, zero, whereas when randomness is
added the fits are much better and yield exponents in the interval $]0,2[$.
In view of the problems justifying such fits in the random case (see below)
this evidence for a softening of the transition is however not to be taken
too seriously.

\subsection{Central charge at the random FP}
\label{Sect:central-charge}

The free energy per site $f^{11}_0(L)$ for the RBPM on long strips of width
$L$ is readily found from Eq.~(\ref{Furstenberg}) applied to the
${\cal T}^{11}$
sector of the TM. We have performed extensive simulations for various
values of $q$ and the randomness strength $R$, though in most cases $R=2$
was found to describe the random FP adequately.

Representative samples of our data are shown in Table \ref{Tab:f}. For each
run a normalised initial vector $|{\rm v}_0 \rangle$ was prepared by
choosing its components randomly, and after discarding the results of the
first 2,000 multiplications by ${\cal T}_i^{11}$ in order to eliminate
transients, data collection was made for each 200 iterations until a strip
of a total length of $m=10^5$ had been built up. For $q>2$ a total of 100
independent runs were made for $1 \le L \le 8$, and 3 runs for
$9 \le L \le 12$, whilst for the Ising model ($q=2$) we were able to make
100 runs for $1 \le L \le 13$ by using the conventional spin basis. Final
results and error bars were extracted by computing the mean and the
standard deviation for the totality of patches of length 200.

It is not {\em a priori} obvious that the Lyapunov exponents found from
Eq.~(\ref{Furstenberg}) are independent of the norm used. The standard norm
in both the spin basis and the connectivity basis is given by
the squareroot of the sum of the squared components, and these two are of
course not identical. To impose the spin basis norm on the connectivity
basis each term in the sum must be weighted by a factor $q^C$, where $C$ is
the number of clusters in the relevant connectivity state. We have checked
the consistency of our results by comparing the first few Lyapunov
exponents obtained from imposing the two different norms on the
connectivity basis, and we find that not only are the results identical but
there is even a complete agreement of the first three significant digits of
the error bars. For $q=2$ we found that the results using the spin basis
and the connectivity basis were consistent, but that the error bars
obtained using the spin basis were slightly smaller.

Our results for the free energies of the random-bond Ising model agree
with, and are more precise than, those of de Queiroz\footnote{Actually they
  differ by a constant, since Ref.~\cite{Queiroz95} defines the Hamiltonian
  as $-\sum_{\langle ij \rangle} K_{ij} \sigma_i \sigma_j$ as opposed to our
  Eq.~(\ref{H}). Since $s_i s_j = 2 \delta_{\sigma_i \sigma_j} - 1$ there
  is a free energy difference of $2 \overline{K_{ij}}$, which for $R=2$
  equals 0.91407.}
\cite{Queiroz95}.

Values of the effective central charge $c'$ can be extracted from
Eq.~(\ref{FSS-c}) by employing various fitting procedures. In spite of the
relatively slow convergence 
of both two-point fits $(L,L+1)$ and straight-line least-squares fits
against $1/L^2$ \cite{Queiroz95}, iterating such fits yields quite good
results in the pure model. When randomness is added this is no longer so,
since rather substantial error bars on the first estimates prevent us from
efficiently iterating the fits.

A better scheme is to include the leading correction to the scaling of
Eq.~(\ref{FSS-c}), which in the pure case has been shown
numerically to take the form \cite{Blote82}
\begin{equation}
  f_0^{11}(L) = f_0(\infty) - \frac{\pi c'}{6 L^2} + \frac{A}{L^4} + \cdots.
  \label{parabolic}
\end{equation}
One then performs either three-point fits $(L,L+1,L+2)$ or parabolic
least-squares fits against $1/L^2$, and because of the much faster
convergence no iteration is needed \cite{Queiroz95}. Although a correction
proportional to $1/L^4$, due to the operator $T \overline{T}$, must
necessarily be present in every system that is conformally invariant
\cite{Cardy-CFT} it can of course not be guaranteed to be the dominant one
in general.

In Table \ref{Tab:c} the results of parabolic fits including the data
points for $L_0 \le L \le L_{\rm max}$ have been shown as a function
of $L_0$. It is seen that $L_0$ must be chosen large enough to justify the
omission of higher terms in the series (\ref{parabolic}), and small enough
to minimise error bars. {}From the special cases of the Ising model and of
the percolation point (see Sect.~\ref{Sect:percolation} below) we concluded
that the choice $L_0 = 3$ is optimal.

Apart from the results shown in Table \ref{Tab:c} we have also
performed some runs for $q=1.5$, finding, as expected from the Harris
criterion \cite{Harris}, no difference between the results
for the pure and the random model.

In the intermediate regime $2 \le q \le 4$ our results compare favourably
to those of the $(q-2)$-expansion, at least up to $q=3$. On the other
hand, it is evident from Fig.~\ref{Fig:c-small} that the difference between
$c$ for the pure model and $c'$ for the random one is of the same order
of magnitude as our error bars, and only near $q=4$, where the expansion is
expected to break down anyway, are our results able to distinguish between
the two different behaviours. Exactly at $q=2$ the randomness is marginal
and logarithmic corrections to the finite-size scaling forms,
Eqs.~(\ref{FSS-x}) and (\ref{FSS-c}), are expected.
Whilst this issue has recently attracted considerable
interest in the case of the critical exponents \cite{Queiroz97} the
corrections to the central charge are much weaker \cite{Cardy86b} and
accordingly our result is consistent with that of the pure Ising model.

In Fig.~\ref{Fig:c-all} we have displayed our results for $c'$ as a
function of $\log_{10}q$ for selected values of $q \in [1.5,64]$. We have
juxtaposed the results for two strengths of the randomness, namely weak
randomness ($R=2$, closed circles on Fig.~\ref{Fig:c-all}) and strong
randomness ($R=10$, open circles). For small values of $q$ both randomness
strenghts give rise to the same $c'$, as witnessed by the overlap of the
$q=4$ data points. However, for larger $q$ the $R=2$ curve flattens out and
grows slower than logarithmically. Sample runs show that the same is true
for larger values of $R$, the difference being that the range of $q$-values
for which the growth is logarithmic is extended as $R$ is increased. This
is illustrated by the $R=10$ curve's staying above, but very close to, the
percolative result $\sim \log q$ (see Sect.~\ref{Sect:percolation} below)
for the whole range of $q$-values shown on the plot. Another way to state
this is that for fixed $q$ and varying $R$, the quantity $c'$ is an
increasing function of $R$ that eventually reaches a plateau as $R$ becomes
large enough. It then appears from Fig.~\ref{Fig:c-all} that for $q \le 64$
the randomness strength $R=10$ is sufficient to reach this plateau.

These findings are interpreted as follows. According to the $(q-2)$-expansion
the randomness strength $R^{\ast}$ corresponding to the random FP is an
increasing function of $q$. Assuming this FP to persist as we enter the
regime $q>4$ (see Fig.~\ref{Fig:phase-diagram}) we now claim that the
monotonicity of $R^{\ast}(q)$ also holds true when the $(q-2)$-expansion
breaks down. {}From the RG flows given in Fig.~\ref{Fig:phase-diagram} we see
that any initial value of $R \in ]1,\infty[$ will eventually flow to the
random FP as the system is viewed on larger and larger length scales.
However, if we start out very far from $R^{\ast}$ the onset of the
asymptotic scaling given by Eq.~(\ref{FSS-c}) may be deferred to much larger
length scales than the strip widths $L$ numerically accessible for our TMs.
We therefore expect poor scaling for strip widths $L \le L_{\rm max}$.
Conversely, if we choose the strength of the randomness as $R \sim R^{\ast}$
the resulting value of $c'$ is expected to be more or less independent of
the precise choice of $R$ and equal to the true value of the central
charge. But in our simulations we find that this is precisely accomplished
by choosing $R$ as an increasing function of $q$.
Further justification for this interpretation
is found from the phenomenological RG treatment in
Sect.~\ref{Sect:phen-renorm} below.

A heuristic argument explaining that the ``effective'' $c'(R,q)$ obtained for
small values of $R$ is less than the ``correct'' value $c'(R^{\ast},q)$
associated with the random FP is readily furnished, at least for large $q$.
Namely, from Zamolodchikov's
$c$-theorem \cite{Zamolodchikov} we know that there exists a function
$c(\{K\})$ of the couplings that decreases along the RG flow and equals the
central charge at the fixed points. As a corollary the curves of constant
$c$ are orthogonal to the RG flow. In particular, for large $q$ where the
RG flow is known from the mapping to the RFIM (see Eqs.~(\ref{rg1}) and
(\ref{rg2})), it is evident from Fig.~\ref{Fig:phase-diagram} that
$c'(R,q)$ is equal to $c'(R^{\ast},q^{\ast})$ for some $q^{\ast} < q$.
Since our numerical results indicate that $c'(R^{\ast},q)$ is an
increasing function of $q$ the proposition follows.

To check our results for $c'$ we have also made 100 independent runs for
each of the strip widths $1 \le L \le 8$
where the random bonds were drawn from the {\em trinary} distribution
\begin{eqnarray}
  P(K) &=& p [\delta(K-K_1) + \delta(K-K_2)] \nonumber \\
       &+& (1-2p)\delta(K-K^{\ast}),
  \label{trinary}
\end{eqnarray}
where $K_1$ and $K_2 = 1000 K_1$ satisfy the criterion (\ref{selfdual}) and
$(\exp K^{\ast} - 1)^2 = q$. Here $p \ll 1$ is the strength of the randomness.
Of course this realisation of the randomness also preserves self-duality, and
hence the model is again at its critical point \cite{Kinzel81}.

Numerical results for $c'$ using trinary randomness are shown in Table
\ref{Tab:c-diluted} and they are consistent with the binary results given
above, again provided that $p$ is increased as we go to larger 
and larger $q$. In particular it is reassuring to verify that we seem to
probe the true random behaviour when $2/p$ (the length scale associated
with this randomness) is comparable to the correlation length of the pure
system (\ref{xi-pure}).

An interesting question is whether the asymptotic value of $c'$ is
approached from above or below when the system is viewed on larger and
larger length scales. For models exhibiting reflection positivity
Zamolodchikov's $c$-theorem \cite{Zamolodchikov} ensures that the
convergence is from above. In particular the condition of positivity holds
true for unitary models, whereas for a random model it may well fail to be
fulfilled. Indeed, in the case of the RBPM a perturbative calculation
\cite{Ludwig-Cardy} suggests that the convergence may be from below in some
cases.

In order to discuss this point the parabolic fits versus $1/L^2$ employed
above are no longer appropriate. Apart from speeding up the rate of
convergence to a point where information about its direction becomes
obliterated due to error bars the inclusion of higher-order corrections to
the finite-size scaling form (\ref{FSS-c}) may have the effect of reversing
this direction. E.g., in the case of the pure Ising model it is found
\cite{Queiroz95} that the estimators obtained from parabolic fits converge
from below, whereas the corresponding linear fits (i.e., without the
$1/L^4$ correction) yield estimators that converge from above in accordance
with the theoretical prediction.

In Table \ref{Tab:c-twopoint} we show the results of such linear
least-squares fits for several values of $q$. The randomness strength $R$
was chosen in accordance with the considerations given above. It appears
that in all cases the finite-size estimators converge towards the
asymptotic values of Table \ref{Tab:c} from above.

We remark that values of $c'$ similar to ours have recently been reported
by Picco \cite{Picco97}. For $q=8$ this author found $c' = 1.45 \pm 0.06$
which agrees with our result of respectively $c' = 1.52 \pm 0.02$ for
binary randomness of strength $R=10$, and $c' = 1.51 \pm 0.04$ for trinary
randomness of strength $p=0.10$. Our observation that $c'$ appears to be an
increasing function of $R$, eventually reaching a plateau as $R$ becomes
large enough, was confirmed by Ref.~\cite{Picco97} that used binary
randomness of strength $R=10$ throughout.
Strong evidence was also given that $c'(q)$ grows
roughly logarithmically with $q$ in the regime $q \in [5,256]$, but a
further discussion of what this implies will be deferred to
Sect.~\ref{Sect:discussion} below.

It is worthwhile to compare the TM algorithm used in Ref.~\cite{Picco97} to
ours. It was found that the number of distinct entries in the pure model TM
in the spin basis is
\begin{equation}
  b_L = \sum_{i_2=1}^2 \sum_{i_3=1}^{m_3} \sum_{i_4=1}^{m_4} \cdots
        \sum_{i_L=1}^{m_L} 1,
  \label{b_L}
\end{equation}
where $m_i = \max(i_2,i_3,\ldots,m_{i-1})$, and $L$ designates the strip
width as usually. Further taking into account the $2^L$ different
realisations of the binary randomness in each strip, recursion relations
between the different elements of the TM were found by computing
a total of $(b_L)^2 2^L$ polynomials. Since this number of polynomials
increases rapidly with $L$ high-precision computations could only be
performed up to $L_{\rm max}=6$. The number of iterations used to determine
$f_0(L=6)$ was similar to ours, whereas more iterations were used for the
smaller strip widths.

Evidently this algorithm also has the advantage that $q$ enters only as a
parameter, thus making accessible any value of $q$ for the
simulations. However, for large $L$ it performs inefficiently, as we will now
show. The numbers $b_L$ of Eq.~(\ref{b_L}) are by no means
unfamiliar. Indeed, they are nothing but the total number of $L$-point
connectivities, including the non-well-nested ones \cite{Jesper2}.
Alternatively they can be viewed as the number of ways that $L$ objects
can be partitioned into indistinguishable parts \cite{Wu97b}.
With $m_{\nu}$ parts of $\nu$ objects each ($\nu = 1,2,\ldots$) 
this can be rewritten as
\begin{equation}
  b_L = \sum_{m_{\nu}=0}^{\infty} {'}
        \prod_{\nu=1}^{\infty}
        \frac{L!}{(\nu!)^{m_{\nu}} \, m_{\nu}!},
\end{equation}
where the primed summation is constrained by the condition
$\sum_{\nu=1}^{\infty} \nu m_{\nu} = L$. {}From this representation the
generating function can be immediately inferred
\begin{equation}
  \exp({\rm e}^t - 1) = \sum_{n=0}^{\infty} \frac{b_n t^n}{n!}.
\end{equation}
Explicit values, found by Taylor expansion of the left-hand side, are shown
in Table \ref{Tab:cn}. Asymptotically the $b_L$ are seen to grow
faster than $L^L$ whereas the well-nested connectivities only grow as
$\sim 4^L$.

\subsection{The percolation limit}
\label{Sect:percolation}

In the case of the binary randomness (\ref{binary}) the percolation
limit is reached by letting $({\rm e}^{K_1} - 1) \rightarrow 0$ and
$({\rm e}^{K_2} - 1) \rightarrow \infty$ whilst maintaining the self-duality
criterion (\ref{selfdual}). The partition function of the random cluster
model is then dominated by one graph only, viz.~the one that covers all of
the strong bonds and none of the weak ones. (Note in particular that the
limits $R \rightarrow \infty$ and $q \rightarrow \infty$ do not commute.)
Expressed in terms of the free energy per site this reads
\begin{equation}
  f_0^{\rm perc} = - \frac{B}{N} \ln ({\rm e}^{K_2} - 1) - \frac{C}{N} \ln q,
\end{equation}
where $B$ is the number of strong bonds and $C$ is the number of clusters
in the dominant graph.

The quenched average over the randomness must be taken on the level of the
free energy. Evidently, with the chosen distribution of the randomness,
$\overline{B} = N$ whence the first term is simply a trivial, albeit
infinite, constant. (Incidentally this is the same constant that was pulled
out in Eq.~(\ref{inf-const}).)
On the other hand, the average number of percolation
clusters is related to a derivative in the {\em pure} $Q$-state Potts model
\cite{Cardy-book}
\begin{equation}
  \overline{C} = \left. \frac{\partial}{\partial Q} \ln Z(Q) \right|_{Q=1},
\end{equation}
thus determining the effective central charge $c'(q)$ at percolation as
\begin{equation}
  c'(q) = \ln q \left. \frac{\partial c(Q)}{\partial Q} \right|_{Q=1}.
  \label{derivative}
\end{equation}
An alternative argument for this relation is furnished by the observation
that the replicated model is simply the Potts model with $q^n$ states;
differentiating $c(q^n)$ with respect to the number of replicas $n$ and
taking the limit $n \rightarrow 0$ one recovers the result (\ref{derivative}).
The central charge of the pure model is given by an expression due to
Kadanoff \cite{Cardy-CFT,Dotsenko84}
\begin{equation}
  c = \frac{(2-3y)(1+y)}{(2-y)},
\end{equation}
where $\sqrt{Q} = 2 \cos (\pi y/2)$ and $0 \le y \le 1$, and taking the
appropriate derivative of this we finally arrive at
\begin{equation}
  c'(q) = \frac{5 \sqrt{3}}{4 \pi} \ln q.
\end{equation}

As described in Sect.~\ref{Sect:T-percolation} the single-bond TMs in the
percolation limit have only one non-zero entry per column, equal to either
$q$, $1$ or $1/q$. Taken together with their projective quality and
Eq.~(\ref{Furstenberg}) for the largest Lyapunov exponent it is clear that
the free energy, and hence the central charge, must be explicitly
proportional to $\ln q$. So it suffices to do the numerics for one value of
$q \neq 1$.
Because of the simple form of these TMs we were able to average
$\tilde{f}_0^{11}(L)$ of Eq.~(\ref{inf-const}) over 100 strips of length
$m=10^5$ for the range $1 \le L \le 19$. Consequently the factor of
proportionality could be determined quite accurately as $0.688 \pm 0.003$,
in excellent agreement with $\frac{5 \sqrt{3}}{4 \pi} \simeq 0.689$.

It is evident from the mapping between bond percolation and the pure $Q=1$
Potts model that the critical exponents of the two models are identical:
$x_T = \frac{5}{4}$ and $x_H = \frac{5}{48}$ \cite{Cardy-CFT}. Since all
correlation functions at percolation can only take the values 0 and 1, it
is also clear that different moments of a given correlation function all
have the same scaling dimension.
Thus, in the notation of Ludwig \cite{Ludwig90},
$x_n = x_1$ for all $n > 1$. The pure model represents the other trivial
extreme case of multiscaling behaviour: $x_n = n x_1$.

\subsection{The cumulant expansion}

The concept of multiscaling is best understood in terms of a simple example
\cite{Derrida84}: The random-bond Ising chain. {}From the reduced Hamiltonian
${\cal H} = - \sum_{i=1}^N K_i s_i s_{i+1}$ the
partition function is easily found as $Z = \prod_{i=1}^N c_i$, where
$c_i = 2 \cosh K_i$. In particular the quenched average
$\overline{Z} = \exp[N \log \overline{c_i}]$ does {\em not} coincide with
the most probable value $Z_{\rm m.p.} = \exp[N \overline{\log c_i}]$.
On the other hand, the reduced free energy is $F = - \sum_{i=1}^N \log c_i$
so that $\overline{F} = F_{\rm m.p.} = - N \overline{\log c_i}$.
The free energy is thus {\em self-averaging}, i.e., it takes on its sample
averaged value with probability unity in the thermodynamic limit. Indeed,
by the central limit theorem, $F$ is normally distributed, it being a sum of
random numbers, whereas $Z$ is a product of random numbers and therefore
{\em log}-normally distributed. Similarly the correlation function
$\langle s_1 s_{R} \rangle = \prod_{i=1}^{R-1} \tanh K_i$ is
non-self-averaging. In particular
$\overline{\langle s_1 s_R \rangle^2} \neq
 \overline{\langle s_1 s_R \rangle}^2$.

In Sect.~\ref{Sect:magnetic} we related the spin-spin correlation function
$G(m)$ on a strip of the RBPM to the free energy in the presence of a seam
of frustrated bonds (or with a ghost site). Taking the logarithm of
Eq.~(\ref{disorder-op}) and exploiting the self-duality of the lattice we have
\begin{equation}
  \Delta f(L) \equiv f_0^{22}(L) - f_0^{11}(L) = \frac{1}{mL} \ln G(m),
\end{equation}
and in the pure system, according to conformal symmetry \cite{Cardy83},
this decays along the strip as $2\pi x_H /L^2$, cfr.~Eq.~(\ref{xh}).
When randomness is present $\Delta f(L)$ is a fluctuating quantity, and
since free energies are supposed to be normally distributed these
fluctuations are ${\cal O}(1/\sqrt{m})$. Consequently $\ln G$ is a
self-averaging quantity and $G$ is not \cite{Derrida84}, exactly as in the
simple example given above.

In the {\em multiscaling} scenario of Ludwig \cite{Ludwig90} different
moments $\overline{G(m)^n}$ scale with dimensions $x_n$ 
which, as opposed to what is the case in the pure model, are not
necessarily linear in $n$.
(In this notation $x_H \equiv x_1$.) For $n_1 > n_2$ we
have $x_{n_1} \ge x_{n_2}$ and $x_{n_1} / n_1 \le x_{n_2} / n_2$
(convexity); pure and percolative behaviour are thus realisations of the
two possible extremes of multiscaling.

Since translational invariance is one of the basic assumptions of conformal
symmetry \cite{Cardy-CFT}, the latter only refers to the averaged quantities
$\overline{G(m)^n}$ and not to the $G(m)^n$ themselves. These averages
cannot be computed directly in a numerical experiment because of the lack
of self-averaging; this can however be circumvented by performing a
cumulant expansion
\begin{equation}
  \ln \overline{G^n} = n \overline{\ln G} +
  \frac{1}{2} n^2 \overline{(\ln G - \overline{\ln G})^2} + \cdots,
  \label{cumulant}
\end{equation}
where each term on the right-hand side {\em is} self-averaging and can be
directly extracted from the statistical fluctuations in $\Delta f(L)$
between the patches of length 200 into which we have divided our stip.

Quite generally for a stochastic variable $X$ we have
\begin{equation}
  \langle \exp X \rangle = \exp \sum_{j=1}^{\infty} \frac{1}{j!} k_j,
\end{equation}
where explicit expressions for the six first cumulants $k_i$ in terms of the
moments $m_i$ of $X$ are given by \cite{Ledermann}
\begin{eqnarray}
  k_1 &=& m_1                                                \nonumber \\
  k_2 &=& m_2 - m_1^2                                        \nonumber \\
  k_3 &=& m_3 - 3 m_2 m_1 + 2 m_1^3                          \nonumber \\
  k_4 &=& m_4 - 4 m_3 m_1 - 3 m_2^2 + 12 m_2 m_1^2 - 6 m_1^4 \nonumber \\
  k_5 &=& m_5 - 5 m_4 m_1 - 10 m_3 m_2 + 20 m_3 m_1^2        \nonumber \\
      &+& 30 m_2^2 m_1 - 60 m_2 m_1^3 + 24 m_1^5             \nonumber \\
  k_6 &=& m_6 - 6 m_5 m_1 - 15 m_4 m_2 + 30 m_4 m_1^2 -
          10 m_3^2                                           \nonumber \\
      &+& 120 m_3 m_2 m_1 - 120 m_3 m_1^3 +
          30 m_2^3 - 270 m_2^2 m_1^2                         \nonumber \\
      &+& 360 m_2 m_1^4 - 120 m_1^6                          \nonumber
\end{eqnarray}
We have computed these six cumulants of $\Delta f(L)$ for various values of
$R$ and $q$, based on 100 independent strips of length $m=10^5$ and width
$1 \le L \le 7$. Sample results for $R=2$ and $q=3,8$ are shown in Table
\ref{Tab:cum}.

For $q=3$ the cumulant expansion converges well. The magnitude of the
higher cumulants decreases very rapidly, especially for $L \ge 3$, and
reliable estimates for the left-hand side of Eq.~(\ref{cumulant}) can be
obtained simply by summing the first 3 or 4 cumulants, at least when $n$ is
not too large. Performing parabolic least-squares fits using Eq.~(\ref{xh})
with an $1/L^4$ correction we thus expect to extract quite accurate values
of $x_n$ at the random FP.

As $q$ increases the convergence is slower. This is witnessed by the $q=8$
results of Table \ref{Tab:cum} decreasing noticeably slower, both for a
definite cumulant as a function of $L$ (vertically) and for a definite $L$
as a function of the cumulant number (horizontally). The approximation of
leaving out the higher cumulants in the sum (\ref{cumulant}) thus becomes
increasingly difficult to justify, and eventually the cumulant
expansion breaks down. This problem is enhanced by the fact that for $q>8$ we
expect a randomness strength of $R=2$ to be insufficient in order to access
the true behaviour at the random FP. We are thus forced to increase $R$,
whence the fluctuations become even more violent and the cumulant expansion
accordingly ill-behaved.

Our results for $x_1$ are shown in Fig.~\ref{Fig:x1}. Since error bars
on the individual cumulants are related to the magnitude of the higher
cumulants the question of how to assign a final error bar to $x_1$ becomes
a delicate one. We have addressed this issue by averaging the estimates for
$x_1$ obtained from various parabolic least-squares fits. More precisely,
the average is calculated from 4 values, namely fits with $L_0 = 3$ or $4$
and including the first 3 or 4 cumulants on the right-hand side of
Eq.~(\ref{cumulant}). The consistency of these 4 values is regarded as a
check of the validity of the expansion.

In particular, for $q=3$ we find $x_1(3) = 0.13467 \pm 0.00013$ which is 10
standard deviations above the value
$x_1^{\rm pure}(3) = \frac{2}{5} \simeq 0.13333$ of the
pure three-state Potts model \cite{Cardy-CFT} and at the same time in perfect
agreement with the result $x_1(3) = 0.13465 + {\cal O}(\epsilon^4)$ of the
$(q-2)$-expansion \cite{Dotsenko95}. The Monte Carlo result
$x_1(3) = 0.1337 \pm 0.0007$ of Picco \cite{Picco96} was not able to
distinguish convincingly between pure and random behaviour.

For $q=4$ our result is $x_1(4) = 0.1396 \pm 0.0005$, in nice agreement
with Picco's preliminary result $x_1(4) \sim 0.139$ \cite{Picco-priv} and
decidedly different from the corresponding pure value of
$x_1^{\rm pure}(4) = \frac{1}{8}$.

As discussed at length in the Introduction a major motivation for this work
was to determine whether the impurity softened transitions for $q>4$ do or
do not have the critical exponents of the pure Ising model. The data of
Fig.~\ref{Fig:x1} clearly show a smooth continuation of the perturbative
results \cite{Ludwig90,Dotsenko95} exhibiting no singularity whatsoever at
$q=4$. Our result $x_1(8) = 0.1415 \pm 0.0036$ is comfortably away from
the pure Ising value and provides a striking piece of evidence for both our
phase diagram and the FP structure of the $(q-2)$-expansion.

All the results quoted for $x_1$ were computed using $R=2$. We have
checked that other values of $R$ yield results consistent herewith,
provided that $R$ is chosen neither too small, in which case the cross-over
length $\xi_X \sim \exp (1/2Aw^2)$ found from Eq.~(\ref{rg1}) becomes too
large for the random FP to be reached, nor too large, in which case the
cumulant expansion breaks down. The same holds true when the random bonds
are drawn from the trinary distribution (\ref{trinary}) with various values
for the dilution parameter $p$.

Because of the positive sign of the second cumulant the values of $x_1$ are
invariably smaller than those one would have obtained without the cumulant
expansion (i.e., using only the first cumulant). The latter, however,
determine a universal exponent $\alpha_0$ that describes the asymptotic
decay of the spin-spin correlation function in a {\em fixed} sample at
criticality. In terms of the multiscaling exponents this reads
\begin{equation}
  \alpha_0 = \left. \frac{{\rm d}x_n}{{\rm d}n} \right|_{n=0}
\end{equation}
Near $q=2$ Ludwig obtained the expansion \cite{Ludwig90}
\begin{equation}
  \alpha_0 = x_1^{\rm pure} + \frac{1}{16} y + {\cal O}(y^2),
  \label{alpha0}
\end{equation}
where $y$ is the RG eigenvalue of the energy operator coupling to the bond
randomness. Our results for small fractional values of $(q-2)$ are in good
agreement with this expression, and if one takes into account the
logarithmic corrections expected exactly at $q=2$ it seems that the
theoretical prediction and the numerical results have a common tangent at
$q=2$. For the physically interesting case of $q=3$ the agreement is not so
good. We believe that this apparent discrepancy would be resolved if the
expansion (\ref{alpha0}) could be carried through to three-loop order as in
the case of $x_1$.

Similar remarks can be made about the higher moments of the spin-spin
correlation function for which we are unable to verify Ludwig's expansion
\cite{Ludwig90}
\begin{equation}
  x_n = n x_1^{\rm pure} - \frac{1}{16} n(n-1) y + {\cal O}(y^2).
\end{equation}
Nevertheless it should be remarked that the fact that the higher cumulants
do not vanish in itself implies multiscaling.

Before concluding this section we should like to give a heuristic argument
that the cumulant expansion breaks down for large $q$. In a replica
formulation we can imagine the central charge $c(n)$ as a function of the
number of replicas $n$. In this notation the central charge of the pure and
the random systems are $c(1)$ and $c'(0)$ respectively, where the prime
denotes a differentiation with respect to $n$. The partition function of
the replicated strip is then
\begin{eqnarray}
  \overline{Z^n} &=& \int \exp(-nmLf)P(f) \,df \nonumber \\
                 &=& \exp \left( -mL \overline{f} + \frac{\pi m c(n)}{6L}
                     - \ldots \right),
\end{eqnarray}
where $P(f)$ is the probability distribution of the free energy.
Differentiating this expression twice with respect to $n$ and taking the
replica limit $n \rightarrow 0$ we infer that the second cumulant of $f$
contains a term that is proportional to $c''(0)$.
The cumulant expansion is thus expected to break down if $c(n)$ has a large
curvature at $n=0$.

For $2 \le q \le 4$ the replicas are weakly coupled, since
$c(1) \simeq c'(0)$ \cite{Ludwig-Cardy}. Hence $c''(0) \ll 1$.
But when $q=4+\epsilon$ the transition of the pure
system goes first order so that the function $c(n)$ starts out with slope
$c'(0) = 1$ and somehow curves down to assume the value $c(1) = 0$.
Consequently $c''(0) = {\cal O}(1)$ and the higher cumulants begin to
contribute significantly to the sum (\ref{cumulant}). Finally, for
$q \gg 4$ we are in the strong coupling regime. We still have $c(1)=0$ and
as our numerical data indicate that $c'(0) \sim \ln q$
it follows that $c''(0) \gg 1$.
This means that the cumulant expansion must break down.

One may speculate whether the transition actually becomes first-order
whenever $q^n > 4$. Clearly this is the case for the pure Potts model
\cite{Baxter73}, but a similar statement is true when $N$ Ising models are
coupled by their local energy density. Namely, in this case an RG analysis
\cite{Cardy96} implies a fluctuation-driven first-order transition whenever
$N>2$, that is to say for $2^N > 4$. If this conjecture is correct one
would then suppose the function $c(n)$ to vanish for $n \ge n_0$, where
$q^{n_0} = 4$. Evidently such a scenario is in accordance with our
observation that $c''(0) \gg 1$ for $q \gg 4$.

\subsection{The thermal exponent}

Because of the rather striking success of the cumulant expansion for $x_1$
one would now expect the thermal exponent $x_T$ to be similarly related to
the fluctuations of $\Delta f_T(L) = f_1^{11}(L) - f_0^{11}(L)$.
Surprisingly, this seems not to be the case. Computing the equivalent of
$\alpha_0$, i.e., using only the first cumulant, we find the following
results for different values of $q$:
$\alpha_0^T(2) = 1.028 \pm 0.001$, $\alpha_0^T(3) =  0.91 \pm 0.01$,
$\alpha_0^T(4) =  0.81 \pm 0.02$ and $\alpha_0^T(8) =  0.65 \pm 0.01$.
As remarked above the results using more cumulants can only be lower.

This is bad news since the quenched correlation length exponent $\nu$ can
be shown quite rigorously to satisfy the bound \cite{Chayes86,Chayes89}
\begin{equation}
  \nu \ge \frac{2}{d},
  \label{Chayes}
\end{equation}
or, in our notation, $x_T \ge 1$. Though the proof of Ref.~\cite{Chayes86}
refers to the divergence of the correlation length as the critical point is
approached, and hence strictly speaking does not apply to the system under
consideration since we work exactly {\em at} the critical point, the RBPM
is among the simplest physical systems for which Eq.~(\ref{Chayes}) is
believed to be valid \cite{Chayes89}. The point is strengthened by noting
that the $(q-2)$-expansion yields $x_T = 1.02 + {\cal O}(\epsilon^3)$
at $q=3$ \cite{Ludwig87}. It is therefore difficult to have confidence in the
cumulant expansion for the thermal exponent, and independent methods of
assessing $x_T$ must be devised.

At this point we note that although the RG equation (\ref{rg3}) seems to
warrant an effective exponent of $x_T^{\rm eff} = 1 - Aw^2$ for $q$
large, this argument is only superficially true. Indeed, near $q = \infty$
the RG flows must extend to infinite $w$ before reaching the random FP, and
consequently an expansion valid for weak randomness is not to be trusted.

The alternative method for finding $x_T$ that comes closest to the spirit
of Refs.~\cite{Chayes86,Chayes89} is that of finite-size scaling off the
critical point. This is discussed at length in the next subsection, and for
the moment we concentrate on less ``obvious'' possibilities.

One of the key points in the construction of the cumulant expansion was the
realisation that the spin-spin correlation function was mapped onto a
surface tension under duality, and hence could be expressed in terms of the
{\em largest} Lyapunov exponent of a TM with twisted boundary
conditions. Reinterpreting the latter as a free energy the self-averaging
property was evident, and the cumulant expansion correspondingly behaved
quite well if the fluctuations were not too large.
It has recently been shown \cite{Jesper2,Wu97b} that under duality four-point
correlation functions are similarly mapped onto (generalised) surface
tensions. Presently these duality relations have only been worked out for
planar graphs, but there is some hope that they may be extended to the case
of cylindrical boundary conditions as well. Taking two of the points as
nearest neighbours on either end of the cylinder we would then recover the
energy-energy correlator, and if the corresponding boundary conditions can
be implemented in the TM $x_T$ follows from a cumulant expansion.

Next, we discuss the method of iterating orthogonal vectors in order to
extract the second Lyapunov exponent \cite{Benettin} in more physical
terms. The energy-energy correlator (Green's function) can be
written as
\begin{equation}
  \langle E(r_1)E(r_2) \rangle =
  \frac{\mbox{\rm Tr} \, E(r_1) E(r_2) \exp(-{\cal H})}
       {\mbox{\rm Tr} \, \exp(-{\cal H})}.
\end{equation}
Now imagine building up the strip by repeated action by the random TMs on
some initial state situated at $r = -\infty$. When we reach $r_1$ the
system is in a state $|{\rm a}_0 \rangle$ on which we act with the
energy operator to define
$|{\rm b}_0 \rangle = E(r_1) |{\rm a}_0 \rangle$.
After $n$ further iterations these states become
\begin{eqnarray}
  |{\rm a}_n \rangle &=& {\cal T}_n \cdots {\cal T}_2 {\cal T}_1
                         |{\rm a}_0 \rangle \nonumber \\
  |{\rm b}_n \rangle &=& {\cal T}_n \cdots {\cal T}_2 {\cal T}_1
                         |{\rm b}_0 \rangle.
\end{eqnarray}
Defining a new state $|{\rm \tilde{b}}_n \rangle$ by orthogonalising
$|{\rm b}_n \rangle$ with respect to $|{\rm a}_n \rangle$
\begin{equation}
  |{\rm \tilde{b}}_n \rangle = |{\rm b}_n \rangle -
  \frac{\langle {\rm a}_n | {\rm b}_n \rangle}
  {\langle {\rm a}_n | {\rm a}_n \rangle} |{\rm a}_n \rangle
\end{equation}
we find that
\begin{eqnarray}
  \frac{\langle {\rm \tilde{b}}_n | {\rm \tilde{b}}_n \rangle}
       {\langle {\rm a}_n | {\rm a}_n \rangle} &=&
  \frac{\langle {\rm b}_n | {\rm b}_n \rangle}
       {\langle {\rm a}_n | {\rm a}_n \rangle} -
  \frac{\langle {\rm a}_n | {\rm b}_n \rangle \,
         \langle {\rm b}_n | {\rm a}_n \rangle}
       {\langle {\rm a}_n | {\rm a}_n \rangle^2} \nonumber \\
  &=&  \langle E(r_1) E(r_2) \rangle -
       \langle E(r_1) \rangle \, \langle E(r_2) \rangle.
  \label{E-connected}
\end{eqnarray}
Thus the process of orthogonalisation corresponds precisely to subtracting
off the disconnected part of the correlation function.

When $n \gg 1$ the states $|{\rm b}_n \rangle$ and
$|{\rm a}_n \rangle$ are almost identical due to contamination and
have a huge norm $\sim \Lambda_0^n$. The idea of orthogonalising them is
therefore numerically extremely unsound. Fortunately a simple calculation
shows that orthogonalising after $n_1$ iterations and then again after
$n-n_1$ further iterations is equivalent to orthogonalising only once, as
above. Hence, by induction, we are allowed to orthogonalise after each
iteration, leaving us with the method of Benettin {\em et al.}
\cite{Benettin}. Similar observations hold true for the higher Lyapunov
spectrum.

At this point an objection may be raised. Since
\begin{equation}
  \langle {\rm a}_n | {\rm a}_n \rangle =
  \langle {\rm a}_0 | {\cal T}_1^{\dagger} {\cal T}_2^{\dagger} \cdots
                      {\cal T}_n^{\dagger} {\cal T}_n \cdots {\cal T}_2
                      {\cal T}_1 | {\rm a}_0 \rangle,
\end{equation}
where the dagger denotes transposition, the correlation function
(\ref{E-connected}) corresponds to a realisation of the randomness that is
always symmetric around the midpoint of $r_1$ and $r_2$.
{}From the above physical picture leading to Eq.~(\ref{E-connected}) it seems
that what we really ought to compute is
\begin{equation}
  \frac{\langle {\rm \tilde{b}}'_n | {\rm \tilde{b}}_n \rangle}
       {\langle {\rm a}'_n | {\rm a}_n \rangle},
  \label{new-correlator}
\end{equation}
where the (transposed) TMs used to obtain the states on the left implement a
different realisation of the randomness than that used to obtain the states
on the right.

Numerically we are now facing the problem of computing the average of
huge numbers that are no longer necessarily positive. As discussed in
connection with Eq.~(\ref{cumulant}) we do not expect the correlation
function to be self-averaging, and because of possible negative values of
Eq.~(\ref{new-correlator}) the subterfuge of averaging its logarithm will
not help us out. Trial runs seem to indicate that for sufficiently small
values of $q$ and $R$ (such as $q=3$ and $R=2$) the matrix elements
appearing in Eq.~(\ref{new-correlator}) computed for the usual
samples of lenght 200 may have either sign, but that their quotient is
invariably positive. The corresponding result for $x_T$ is roughly equal to
that obtained from the cumulant expansion. Unfortunately, for larger values
of $q$ or $R$ rare events of negative quotients begin to occur, and any
attempt of averaging Eq.~(\ref{new-correlator}) without resorting to
logaritms is hampered by such large fluctuations as to render the results
insufficiently accurate at the very best. Computations along these lines,
though physically appealing, must therefore be abandoned on numerical grounds.

Yet another possibility of determining at least an approximate value of
$x_T$ is through the conformal sum rule \cite{sumrule} that for an $n$-fold
replicated system reads
\begin{equation}
  \frac{c(n)}{12} = \frac{\sum_i d_i(n) x_i {\rm e}^{-2\pi x_i}}
                         {1 + \sum_i d_i(n) {\rm e}^{-2\pi x_i}},
\end{equation}
where the sum runs over all operators in the theory, including the
descendants of the Verma module with their appropriate multiplicities,
and $d_i(n)$ are the multiplicities pertaining to the permutational symmetry
of the $n$ replicas and the $q$ Potts states. For the magnetic operator
$d_i(n) = n(q-1)$, since there are $(q-1)$ independent order parameters,
and in the case of the energy operator $d_i(n) = n$.
In the pure system this yields quite accurate estimates for $x_T$ if the
exact values of $c(1)$ and $x_H$ are inserted along with the first descendant
of the latter. Differentiating and going to the replica limit we find that for
a random system
\begin{equation}
  \frac{c'(0)}{12} = x_1 (q-1) {\rm e}^{-2\pi x_1} -
                     \frac{x_2}{2}(q-1)^2 {\rm e}^{-2\pi x_2} + \cdots,
\end{equation}
so that for values of $x_1$, $x_2$ and $x_T$ near those of the Ising model the
term with $x_T$ enters only as a small correction. Consequently, at the
very best only $x_2$ can be determined with some confidence from our
previous results for $c'(0)$ and $x_1$. Its value appears to be consistent
with that obtained from the cumulant expansion.

Finally we should like to mention that preliminary studies of exact partition
function zeros for small $L \times L$ lattices with quenched bond
randomness hints at an interesting new method of estimating $x_T$.
Although the different realisations of the quenched bond randomness in
general lead to a considerable scatter in the positions of such zeros, it
turns out that the zeros that are closest to the real axis only exhibit a
very weak dependence on the realisation. But these zeros are precisely
those that fix $x_T$ through their finite-size scaling. Results along these
lines, both for zeros of the Lee-Yang and of the Fisher type, will be
published elsewhere \cite{PartFunc}.

\subsection{Phenomenological renormalisation}
\label{Sect:phen-renorm}

In view of the difficulties encountered in our attempts to extract $x_T$
directly at the
critical point we turn our attention to the method of {\em phenomenological
renormalisation} \cite{Nightingale76}, which is closer in spirit to the
ideas that lead to the bound (\ref{Chayes}).

The magnetic correlation length can be found from the TM spectra through
\begin{equation}
  \xi(L,T)^{-1} = \ln \left( \frac{\Lambda_0^{11}}{\Lambda_0^{22}} \right)
                = L (f_0^{22} - f_0^{11}),
\end{equation}
and we note that this quantity would be self-averaging in the random case.
Motivated by the form $\xi \sim (T-T_{\rm c})^{-\nu}$ of the divergence of the
correlation length in the infinite system we make the finite-size scaling
ansatz
\begin{equation}
  \xi(L,t) = L \phi \big( (T-T_{\rm c}) L^{1/\nu} \big).
  \label{FSS-ansatz}
\end{equation}
For pure systems, then, one traditionally scans through the vicinity of
$T_{\rm c}$ to find an effective $T_{\rm c}(L)$ as the solution of
\begin{equation}
  \frac{\xi(L,T_{\rm c}(L))}{L} = \frac{\xi(L-1,T_{\rm c}(L))}{L-1},
\end{equation}
and computes an approximant $\nu(L)$ from
\begin{equation}
  1 + \frac{1}{\nu(L)} = \left.
  \frac{\ln \big( \mu(L,T) / \mu(L-1,T) \big)}
       {\ln \big( L / (L-1) \big)} \right|_{T=T_{\rm c}(L)},
  \label{nu(L)}
\end{equation}
where the derivatives
\begin{equation}
  \mu(L,T) \equiv \frac{\partial \xi(L,T)}{\partial T} =
  L^{1+\frac{1}{\nu}} \phi' \big( (T-T_{\rm c}) L^{1/\nu} \big)
  \label{mu(L)}
\end{equation}
are found by numerical differentiation.
As $L \rightarrow \infty$ we have $T_{\rm c}(L) \rightarrow T_{\rm c}$ and
$\nu(L) \rightarrow \nu$.

In the random case the extracted values of $\xi(L,T)$ are hampered by
substatial error bars, and the method just outlined becomes by far too
inefficient. Fortunately the very costly idea of scanning for $T_{\rm c}(L)$
can be discarded, since the exact $T_{\rm c}$ of the RBPM is known from
Eq.~(\ref{selfdual}). Consequently the derivatives (\ref{mu(L)}) and the
approximants (\ref{nu(L)}) are evaluated at the exact $T_{\rm c}$, whence
the only remaining source of errors is that of statistical fluctuations
over the different realisations of the randomness.

Na\"{\i}vely one would now find the derivative (\ref{mu(L)}) by subtraction of
the free energies evaluated at $T = T_{\rm c} (1 \pm \epsilon)$, where
$\epsilon \ll 1$. Although this method yields reasonable results for
$\epsilon \sim 10^{-2}$ it is way too inaccurate, since it involves the
subtraction of almost identical quantities (with error bars). A superior
strategy is to divide the strip into patches of length 200, calculate
{\em exact}\footnote{Since we are now faced with differentiating a quantity
  that is known with full machine precision ($10^{-16}$) we can concentrate
  on minimising the rounding and truncation errors. This is accomplished by
  choosing $\epsilon = 10^{-5}$. See Sect.~5.7 of:
  W.~H.~Press {\em et al.}, {\em Numerical Recipes in C}, second edition
  (Cambridge University Press, Cambridge, 1992).}
values of $\mu(L)$ for each of those, and
finally average over the totality of such patches. In this way one exploits
the fact that
$\xi \big( L,T_{\rm c}(1 - \epsilon) \big)$ and
$\xi \big( L,T_{\rm c}(1 + \epsilon) \big)$ are strongly correlated when
the realisation of the randomness is kept fixed. In practice we found that
this trick lead to a reduction of the error bars with a factor $\sim 120$.

Sample results obtained by using these prescriptions are shown in Table
\ref{Tab:phen-renorm}. It is seen that although the convergence is still too
slow for reliable extrapolations to the limit of an infinite system to be
made, the conflict with the bound (\ref{Chayes}) appears to be resolved.

We have found that the convergence of the estimators $\nu(L)$ can be
significantly sped up if one performs the numerical differentiation
(\ref{mu(L)}) by going perpendicularly off the self-duality criterion
in $(K_1,K_2)$ space instead of maintaining the condition
$K_2 = R K_1$. Indeed, one may imagine that there is another exponent
associated with motion {\em along} the critical surface, and maintaining
$K_2 = R K_1$ one would then measure an admixture of this spurious
exponent, in particular for large $R$. A simple calculation using
Eq.~(\ref{selfdual}) shows that one should then evaluate $\xi(L,K_1',K_2')$
at
\begin{eqnarray}
  K_1' &=& K_1 \left(1 \pm \epsilon
           \frac{R {\rm e}^{K_1}}{q {\rm e}^{K_2}} ({\rm e}^{K_2}-1)^2
           \right) \nonumber \\
  K_2' &=& K_2 (1 \pm \epsilon).
  \label{perpendicular}
\end{eqnarray}
The sample results shown in Table \ref{Tab:perpendicular} exhibit a
conspicuous improvement over those of Table
\ref{Tab:phen-renorm}. Not only is the convergence faster, but it is even
seen that the estimators $\nu(L)$ form a monotonically increasing sequence
for low values of $R$ and a monotonically decreasing one for high $R$. The
extrapolated $\nu$ is pinched between those two sequences and consequently
quite accurately determined.

Plots of the estimators $\nu(L)$ for $3 \le L \le 7$ and several values of
$R$ are shown in Figs.~\ref{Fig:xtq8} and \ref{Fig:xtq64} for $q=8$ and
$q=64$ respectively. It is seen that the curves for $\nu(L)$ and $\nu(L-1)$
intersect at a unique value of $R$, seemingly converging quite fast to a
definite value $R^{\ast}$ as $L$ increases. We interpret $R^{\ast}$ as the
randomness strength at the critical FP and the corresponding value of $\nu$
as the correct critical exponent. It is tempting to conjecture that the
curves $\nu(L)$ approach $\nu$ on the entire interval
$R \in ]1,\infty[$ as $L \rightarrow \infty$. {}From the graphs it seems
that the convergence is faster for large $q$.

The values of $\nu$ and $R^{\ast}$ corresponding to this scenario are shown
in Table \ref{Tab:nu}. In accordance with the phase diagram
(Fig.~\ref{Fig:phase-diagram}) $R^{\ast}$ is a slowly, supposedly
logarithmically\footnote{This supposition constitutes the simplest
  possibility allowed by the phase diagram of Fig.~\ref{Fig:phase-diagram}
  in which $R^{\ast} \rightarrow \infty$ as $\ln q \rightarrow \infty$.}
, increasing function of $q$. For $q=2$ the deviation from 
$\nu=1$ can be ascribed to logarithmic corrections \cite{Queiroz97}, and
for $q=3$ our result for $\nu$ is in agreement with the $(q-2)$-expansion
\cite{Ludwig87} though the possibility of replica symmetry breaking cannot
be ruled out \cite{Dotsenko96}. Also for $q>4$ are our values for $\nu$
numerically consistent with unity, indicating that, unlike what is the case
for the magnetic expononent, the thermal one displays only a weak
$q$-dependence.

{}From Figs.~\ref{Fig:xtq8} and \ref{Fig:xtq64} a remarkable feature about the
pure system ($R=1$) is apparent. For $q=8$ the estimators $\nu(L)$ seem to
converge to $\nu = \frac{1}{2}$ whilst for $q=64$ the extrapolated value is
$\nu \simeq 0$. The former value is hardly surprising since, as we also
remarked above, a first-order transition is expected to exhibit scaling
with trivial effective exponents (in this case: $x_T = 0$). On the other
hand, $\nu \simeq 0$ for $q=64$ has to do with the length scales of the
system. Namely, from the asymptotic behaviour of the correlation length of
the pure system \cite{Buffenoir}
\begin{equation}
  \xi \sim \frac{2}{\ln q} \mbox{\rm ~as } q \rightarrow \infty
\end{equation}
we infer that $\xi \sim 1 \ll L$ at the transition point of the $q=64$ system.
But away from the transition point we also expect $\xi = {\cal O}(1)$,
since the lattice spacing (unity) is the least length scale in the
system. After all there is a ferromagnetic interaction between nearest
neighbour spins. We thus conclude that $\xi$ is roughly temperature
independent in this case. In order for this to be consistent with the
asymptotic behaviour of the finite-size scaling ansatz (\ref{FSS-ansatz})
\begin{equation}
  \phi(x) \sim x^{-\nu} \mbox{\rm ~for } x \ll 1
\end{equation}
we must then have $\nu \simeq 0$. This is to be contrasted to the case of
$q=8$ where $\xi \gg L$ so that we clearly ``see'' the phase transitions in
our strips of width $L$.

Very recently the bound (\ref{Chayes}) was challenged by P\'{a}zm\'{a}ndi
{\em et al.}~\cite{Pazmandi}. These authors claimed that the standard
method of averaging over the disorder in finite-size (FS) systems
introduces a new diverging length scale into the problem, whence the
resulting $\nu_{\rm FS}$ may be unrelated to the true exponent $\nu$
governing the divergence of the correlation length in the infinite
system. In particular $\nu$ can be less than $\frac{2}{d}$, and if this is
the case the standard method is liable to yield exactly
$\nu_{\rm FS} = \frac{2}{d}$.
Ref.~\cite{Pazmandi} then went on to suggest a noise reduction procedure
that professedly would allow one to access the true $\nu$. For each
realisation of the binary randomness (\ref{binary}) used in the disorder
average there will be a fraction $p$ of weak bonds ($K_1$). The noise due to
the fluctuations of $p$ around its average value $\overline{p} = \frac{1}{2}$ 
can then be reduced by adjusting the couplings $(K_1,K_2)$ for that
particular realisation to the values they would assume at the critical
point of an infinite system with a (mean) fraction of weak bonds equal to $p$.

To implement this we are faced with the task of finding the two-dimensional
critical surface in the space $(K_1,K_2,p)$ using our knowledge of its
one-dimensional intersection with the plane $p=\frac{1}{2}$,
viz.~Eq.~(\ref{selfdual}). Let the fraction of weak bonds in a particular
realisation be
\begin{equation}
  p = \frac{1}{2} (1 + \epsilon_p),
\end{equation}
where $\epsilon_p \ll 1$. The symmetry $p \leftrightarrow 1-p$ ensures
that, to first order in $\epsilon_p$, we must still go perpendicularly off 
the self-dual curve in the $(K_1,K_2)$ subspace, as in
Eq.~(\ref{perpendicular}). Since an increase in the number of weak bonds
must be offset by an increase of the $K$'s in order to keep the coupling to
the energy density constant, the correct prescription is
\begin{eqnarray}
  K_1 &\rightarrow& K_1 \left(1 + \epsilon_K
       \frac{R {\rm e}^{K_1}}{q {\rm e}^{K_2}} ({\rm e}^{K_2}-1)^2 \right)
       \nonumber \\
  K_2 &\rightarrow& K_2 (1 + \epsilon_K)
\end{eqnarray}
for some $\epsilon_K > 0$.
Demanding that the combined change in $p$ and in $(K_1,K_2)$ must leave the
coupling to the energy density invariant furnishes a relation between
$\epsilon_p$ and $\epsilon_K$
\begin{equation}
  \epsilon_K = \epsilon_p \frac{K_2 - K_1}
  {K_1 \frac{R {\rm e}^K_1}{q {\rm e}^{K_2}} ({\rm e}^{K_2}-1)^2 + K_2}
\end{equation}
and the derivatives (\ref{mu(L)}) are now evaluated at these values of the
parameters by going perpendicularly off the critical surface. To first
order, of course, Eq.~(\ref{perpendicular}) still gives the correct way
of doing so.

Our confidence in the results of Table \ref{Tab:nu} is increased by
observing that the implementation of this novel averaging procedure does
{\em not} alter our results. Indeed, trial runs for $q=8$, where the
discrepancy between the $x_T$ extracted from the Lyapunov spectrum and
phenomenological RG respectively is large, render the values of the
estimators $\mu(L)$ unchanged within (small) error bars.
It is thus concluded that even though our results for $\nu$ are
conspicuously close to satisfying the bound (\ref{Chayes}) with equality,
this is not due to an artifact in the averaging procedure.

\subsection{The higher Lyapunov spectrum}

Although the second Lyapunov exponent of ${\cal T}^{11}$ fails to yield
the thermal scaling dimension $x_T$ in the standard way it is hard to
believe that the
Lyapunov spectrum is not in some way related to the operator content of the
CFT underlying the RBPM. In the case of the pure three-state Potts model,
for example, it is well known \cite{Cardy-CFT} that the first five gaps of
the $Z_2$-even sector are related to the energy density $\epsilon$,
its first descendants $L_{-1}\epsilon$ and $\overline{L}_{-1}\epsilon$,
the stress tensor $T$ and its conjugate $\overline{T}$. To wit, the scaling
dimensions of these operators can be found from the gaps through
Eq.~(\ref{FSS-x}), and we have verified this using our connectivity basis TMs.

In view of the bound (\ref{Chayes}) it is problematic to associate the
first gap with the energy density in the random case, but it is
nevertheless a beguiling question whether such concepts as descendants and
the stress tensor are preserved by the randomness. To investigate this
issue we have computed the first few Lyapunov exponents of ${\cal T}^{11}$
for $1 \le L \le 8$, averaging over 100 runs as usual. The scaling
dimensions corresponding to the first five gaps for $q=3,4,5$ and $R=2$ are
shown in Table \ref{Tab:Lyapunov}. Self-averaging was ensured by utilising
the cumulant expansion, and parabolic least-squares fits included the first
three cumulants.

It is quite remarkable that even if the scaling dimension corresponding to
the first gap may not be equal to $x_T$ our data give strong reasons to
believe that it has a descendant, and that this descendant has the expected
degeneracy of two. And even though the scaling dimensions in general depend
on $q$ those corresponding to the fourth and the fifth gaps are constant within
error bars and very close to 2, as is expected for the stress tensor of a
conformally invariant system \cite{Cardy-CFT}. Preliminary data for even
higher Lyapunov 
exponents seem to hint at descendants at level two, but since we have found
that in the pure system higher and higher eigenvalues require larger and
larger system sizes before the asymptotic scaling form (\ref{FSS-x}) is
valid, massive computations are needed to establish reliable results
for all but the first few scaling dimensions.

Another interesting feature of our data for the higher Lyapunov exponents
is that the Harris criterion seems to be valid in a very complete
sense. Namely, trial runs for $q=1$ seem to indicate that although
individual cumulants exhibit a pronounced dependence of $R$, their sum is
virtually independent of the strength of the randomness in the whole range
$R \in [1,2]$.
It thus appears that all exponents $x_i$ that we can extract numerically from
the Lyapunov spectrum, using Eq.~(\ref{FSS-x}) and the cumulant expansion,
obey the Harris criterion. Since the connection between these exponents and
the scaling dimensions of the underlying CFT is not completely known
(witness $x_T$) this may well turn out to be a non-trivial observation.

\section{Discussion and outlook}
\label{Sect:discussion}

In a recent paper by Picco \cite{Picco97} it has been suggested that for
$q=2^p$ the effective central charge at the random FP is
$c' = \frac{p}{2}$, and that this class of models thus behaves as $p$
decoupled Ising models. Even without referring to our values of the
magnetic exponent we should like to point out that all the data show is
that $c' \propto \ln q$ with a constant of proportionality that is very
close to $\frac{1}{2 \ln 2} \simeq 0.721$. But this constant is {\em also}
very close to that of the percolation point,
viz.~$\frac{5 \sqrt{3}}{4 \pi} \simeq 0.689$. Indeed, these two numbers
differ by less than 5\%, and since our error bars and those of Picco
are in the 2\% and the 4\% range respectively, there is no irrefutable way
of numerically distinguishing between percolative, Ising-like or indeed
some other,
presently unknown, behaviour of the central charge. A similar observation
is valid for $2 \le q \le 4$ where our numerical data as displayed in
Fig.~\ref{Fig:c-small} are compatible, within error bars, with both the
values at the pure and the random FP (but not, in this case, with those at the
percolation point).

On the other hand, our results for the magnetic exponent should leave no
doubt that the correct CFT describing the RBPM cannot be that of a number
of decoupled Ising models. In particular, the non-Ising value at $q=8$ is
in sharp disagreement with the Monte Carlo results of Ref.~\cite{Chen95}.
One possible explanation of this discrepancy is that these authors define
a non-standard order parameter through
\begin{equation}
  \rho = L'^{-d} \langle\max (N_1,N_2,\ldots,N_q)\rangle,
\end{equation}
where $N_i$ is the number of spins in state $i$, which is
related to our local order
parameter defined in Eq.~(\ref{orderparameter}) by $N_i=\sum_r(M_i(r)+q^{-1})$.
The site label $r$ runs over
a hypercube of side $L'$ with $24 \le L' \le 84$. 
Note that $\rho$ may also be written as 
\begin{equation}
\rho= L'^{-d}
  \lim_{k \rightarrow \infty} \left( \sum_{i=1}^q \langle N_i^k \rangle
\right)^{1/k}.
\end{equation}
Expressed in terms of the local order parameter, $\langle N_i^k \rangle$
gives a sum of terms each of the form
\begin{equation}
\sum_{r_j}\langle M_i(r_1)^{k_1}M_i(r_2)^{k_2}\ldots M_i(r_n)^{k_n}\rangle,
\end{equation}
where $k_1+k_2+\cdots=k$. As $k\to\infty$ at fixed $L'$, it is clear
that at least some of the $k_j$ must grow large. In the pure system,
this should not matter, since each term
will scale in the same manner. But when multiscaling is present, the
scaling behaviour of the $k_j$ power of the local order parameter
may be different.
Indeed, in the
limit of $k \rightarrow \infty$ one would expect $\rho$ to scale with
dimension $\lim_{k \rightarrow \infty} x_k / k$, which is {\em less} than
$x_1$ by convexity.

Another criticism of Ref.~\cite{Chen95} is that the realisations of the
binary randomness considered were confined to those for which the number
of strong and weak bonds in each of the two lattice directions were
equal. Though this restriction is clearly innocuous in the limit
$L' \rightarrow \infty$ this may not be so as far as the finite-size scaling
is concerned. {}From trial runs where similar restrictions were imposed to
the bond distributions of the TMs we found that seemingly harmless noise
reductions schemes can influence the output substantially.

Finally, the mapping to the RFIM \cite{Jesper3} illustrated that for large
$q$ typical configurations consist of large clusters of spins in the same
Potts state, separated by domain walls. Whilst our very long strips are
guaranteed to accomodate large regions in which all $q$ values of the order
parameters are realised, it is not clear that this should be the case in
the much smaller square geometries of Ref.~\cite{Chen95}. Indeed it seems
likely that one would find Ising exponents if the geometry under
consideration typically can accomodate at most two different large
clusters.

We now turn our attention to the thermal exponent. If the phenomenological
RG scheme is to be trusted the values of $x_T$ only exhibit a weak
dependence on $q$, although the $(q-2)$-expansion gives us reason to
believe that there is some variation \cite{Ludwig87}. It is
interesting that $x_T$ stays so close to unity even at very high $q$, but
presently we do not have any arguments to explain this
finding. Unfortunately the method employed was unable to resolve
the slight deviations from unity, and it is indeed a challenge to future
research to find more accurate ways of assessing $x_T$ for disordered
systems. Our results on the higher Lyapunov spectrum are nothing if not
intriguing, and we believe that a great effort must be made to understand
why the first gap in the spectrum fails to be related to $x_T$ in the
standard way, even though the higher gaps show clear indications of a
conformal field theory underlying the RBPM.

A very interesting issue to be addressed by future research is that of the
dynamical universality class of the RBPM. In particular it would be
interesting to see whether the dynamic critical exponent $z$ does or does
not agree with the Ising value of $z \approx 2$, or whether, in analogy
with the RFIM, there is logarithmically slow dynamics due to the pinning
of domain walls by impurities.

Other types of randomness are also of interest to the question whether a
first-order phase transition is softened due to impurities. In this paper
we have studied the effect of quenched bond randomness in a flat, regular
lattice. A somewhat different scenario is obtained by investigating the
pure $q$-state Potts model on lattices with quenched {\em connectivity}
disorder. In Ref.~\cite{Janke} MC simulations of the $q=8$ model on
two-dimensional Poissonian random lattices (Voronoi tessellations) with
toroidal topology unambiguously showed that the first-order nature of the
transition was not modified.

An argument why this must be so, at least for large $q$, can readily be
given. For simplicity consider the model on the dual Delaunay random
lattice, which per construction is a triangulation of space \cite{Okabe}.
As on the regular lattice, at large $q$ there are only two important states
in the equivalent random cluster model: the empty lattice, which
contributes a term $q^{N_{\rm vertices}}$ to the partition function, and the
full lattice, contributing a factor $u^{N_{\rm edges}}$,
where $u=e^K - 1$. Since for any triangulation
$2 N_{\rm edges} = 3 N_{\rm vertices}$, the transition occurs when
$u \sim q^{2/3}$ for {\em any} geometry. If we now consider that part of
the lattice within a large hypercube of side $L$, the fluctuations in the
difference of the energies of these two states inside this region will come
solely from the edges which penetrate the boundary. On average, the
difference in the energies of these two states will still be zero, but
there will be fluctuations of the order of the square root of the number
of bonds which penetrate the boundary, which will therefore be
${\cal O}(L^{(d-1)/2})$. These are very much smaller than the analogous
fluctuations which are present when random bonds are added: these are
${\cal O}(L^{d/2})$, which leads to the Imry-Ma type of argument
\cite{Imry-Ma,Hui-Berker,Aizenman-Wehr}. For $d=2$ Voronoi tessellations
the fluctuations are thus always smaller than the domain wall energy
${\cal O}(L^{d-1})$, and we conclude that such randomness is strongly
irrelevant (at least for large $q$), in agreement with the results of
Ref.~\cite{Janke}.

Yet another kind of randomness is obtained by studying the Potts model on
quenched random gravity graphs, for which MC simulations for
$q=10$ have provided strong evidence for a softening scenario similar to
ours \cite{Baillie}. However, in this case the {\em curvature} is random
and hence when the lattice is embedded in the plane, it is fractal.
Although the argument about compensation of the bulk energies when
$u \sim q^{2/3}$ works for any triangulation, the number of boundary edges
may well scale in a different manner for these lattices. Whilst it would be
interesting to study this in detail, it is clear that this problem is quite
different from ours, and neither our arguments nor those of
Refs.~\cite{Hui-Berker,Aizenman-Wehr} can be directly applied.

\begin{acknowledgements}

Stimulating discussions with J.~Chalker, E.~Domany, M.~Kardar, D.~P.~Landau,
A.~Ludwig and S.~de Queiroz are gratefully acknowledged. 
We also thank the staff at the Institute for Theoretical Physics at Santa
Barbara, where part of this work was done, for its warm hospitality.
This research was supported in part by the Engineering and Physical Sciences
Research Council under Grant GR/J78327, and by the National Science Foundation
under Grant PHY94-07194.

\end{acknowledgements}

\end{multicols}

\newpage

\begin{figure}
  \caption{Schematic phase diagram in the critical surface for $d>2$. $q$
    increases to the left and $w$ is the disorder strength, with $P_1P_2$
    being the percolation limit. RG flows are indicated. The latent heat is
    non-vanishing within the shaded region, and elsewhere the transition is
    continuous, controlled by the line of fixed points $P_1q_1$.
    As $d \rightarrow 2$ the shaded region collapses to a line $q_2O$ of
    first-order transitions in the pure system.}
  \label{Fig:phase-diagram}
\end{figure}

\begin{figure}
  \caption{Plots of $\lambda(L)$, normalised to $\lambda(1)=1$, showing
    that bond randomness renders the phase transition second order. The
    random systems have $R \equiv K_2/K_1 = 2$.}
  \label{Fig:12}
\end{figure}

\begin{figure}
  \caption{The effective central charge $c'$ as a function of $q$ for
    $2 \le q \le 4$. The perturbative results by Ludwig and Cardy [14]
% Should be \cite{Ludwig-Cardy}, but RevTeX macro doesn't accept \cite here.
    do not differ appreciately within the range of
    $q$-values where the expansion is supposed to be valid. Accordingly the
    numerical data are unable to distinguish between pure and non-trivial
    random behaviour. They are also quite close to, but distinguishable
    from, the percolation point values.}
  \label{Fig:c-small}
\end{figure}

\begin{figure}
  \caption{Effective central charge as a function of $\log_{10}q$ for
    $1.5 \le q \le 64$. For large $q$ the data for $R=10$ are supposed to
    represent the true behaviour at the random fixed point, as argued in
    the text.}
  \label{Fig:c-all}
\end{figure}

\begin{figure}
  \caption{Magnetic exponent $x_1 = \beta / \nu$ for $R=2$ as obtained from
    the cumulant expansion. $x_1$ is an increasing function of $q$,
    continuously connecting onto the perturbative results near $q=2$. For
    $q>8$ the cumulant expansion begins to break down.}
  \label{Fig:x1}
\end{figure}

\begin{figure}
  \caption{Estimants $\nu(L)$ for the thermal exponent at $q=8$ as obtained
    from phenomenological renormalisation applied to strips of width $L$
    and $L-1$. In the pure system ($R \rightarrow 1$, see rightmost inset)
    the estimants converges to $\nu(\infty) = \frac{1}{2}$. Curves for
    neighbouring system sizes intersect at values of $\nu$ and $R$ that
    converge to those at the random fixed point as $L \rightarrow
    \infty$. In this case $\nu = 1.01 \pm 0.02$ and $R^{\ast} \approx 9$
    (see left inset). Error bars are no larger than the size of the symbols.}
  \label{Fig:xtq8}
\end{figure}

\begin{figure}
  \caption{Phenomenological renormalisation at $q=64$. The curves intersect
    at larger angles than before, allowing for a rather accurate
    determination $\nu = 1.02 \pm 0.03$ in spite of the large fluctuations.
    Error bars are comparable to the size of the symbols. From the rightmost
    inset it is seen that $\nu \rightarrow 0$ in the pure systems, as
    explained in the text. The left inset is a magnification of the region
    around $R^{\ast} \approx 10$.}
  \label{Fig:xtq64}
\end{figure}

\newpage

\begin{table}
 \begin{tabular}{rrrrrr}
   $L$ &   $c_L$ &     $d_L$ & $d_L-c_L$ &   $L c_L$ &     $b_L$ \\ \hline
     1 &       1 &         2 &         1 &         1 &         1 \\
     2 &       2 &         5 &         3 &         4 &         2 \\
     3 &       5 &        15 &        10 &        15 &         5 \\
     4 &      14 &        51 &        37 &        56 &        15 \\
     5 &      42 &       188 &       146 &       210 &        52 \\
     6 &     132 &       731 &       599 &       792 &       203 \\
     7 &     429 &     2,950 &     2,521 &     3,003 &       877 \\
     8 &   1,430 &    12,235 &    10,805 &    11,440 &     4,140 \\
     9 &   4,862 &    51,822 &    46,960 &    43,758 &    21,147 \\
    10 &  16,796 &   223,191 &   206,395 &   167,960 &   115,975 \\
    11 &  58,786 &   974,427 &   915,641 &   646,646 &   678,570 \\
    12 & 208,012 & 4,302,645 & 4,094,633 & 2,496,144 & 4,213,597 \\
 \end{tabular}
 \caption{The number of connectivity states for a Potts model transfer
          matrix of width $L$ with ($d_L$) and without ($c_L$) an external
          magnetic field. Also shown is the size of the magnetic sector
          when using a ghost site ($d_L-c_L$) and a seam ($Lc_L$).
          For large strip widths the seam is advantageous. The number $c_L$
          of well-nested $L$-point connectivities should be compared to the
          total number of $L$-point connectivities $b_L$ which increases
          {\em faster} than exponentially as a function of $L$.}
 \label{Tab:cn}
\end{table}

\begin{table}
 \begin{tabular}{rllllll}
   $L$ & $q=2$             &  $q=3$         &  $q=4$
       & $q=8$             &  $q=16$        &  $q=64$        \\ \hline
     1 & 2.17460   (12)    &  2.62881 (13)  &  2.96193 (13)
       & 3.80035   (16)    &  4.68198 (18)  &  6.54635 (24)  \\
     2 & 1.95329   (8)     &  2.26650 (9)   &  2.49558 (9)
       & 3.06980   (10)    &  3.67393 (11)  &  4.95619 (14)  \\
     3 & 1.90971   (7)     &  2.19534 (7)   &  2.40405 (7)
       & 2.92819   (8)     &  3.48241 (9)   &  4.67423 (11)  \\
     4 & 1.89550   (6)     &  2.17203 (6)   &  2.37431 (6)
       & 2.88328   (7)     &  3.42387 (8)   &  4.59557 (10)  \\
     5 & 1.88895   (5)     &  2.16182 (6)   &  2.36126 (6)
       & 2.86392   (6)     &  3.39934 (7)   &  4.56442 (9)   \\
     6 & 1.88568   (5)     &  2.15649 (5)   &  2.35449 (5)
       & 2.85395   (6)     &  3.38683 (6)   &  4.54893 (8)   \\
     7 & 1.88377   (4)     &  2.15328 (5)   &  2.35040 (5)
       & 2.84798   (5)     &  3.37948 (6)   &  4.53974 (7)   \\
     8 & 1.88250   (4)     &  2.15113 (4)   &  2.34782 (4)
       & 2.84424   (5)     &  3.37479 (5)   &  4.53394 (7)   \\
     9 & 1.88164   (4)     &  2.14993 (24)  &  2.34624 (25)
       & 2.84172   (11)    &  3.37186 (31)  &  4.53017 (39)  \\
    10 & 1.88098   (4)     &  2.14858 (23)  &  2.34504 (23)
       & 2.84011   (26)    &  3.36918 (29)  &  4.52653 (25)  \\
    11 & 1.88048   (4)     &  2.14804 (22)  &  2.34386 (22)
       & 2.83851   (24)    &  3.36768 (29)  &  4.52465 (34)  \\
    12 & 1.88017   (3)     &  2.14744 (20)  &  2.34314 (21)
       & 2.83765   (24)    &  3.36639 (26)  &  4.52316 (34)  \\
    13 & 1.87991   (3)     &                &
       &                   &                &                \\
 \end{tabular}
 \caption{Critical free energies per site, $-f_0^{11}(L)$, for $R=2$ and
   various values of $q$. The figures in parentheses indicate the error bar
   on the last quoted digits.}
 \label{Tab:f}
\end{table}

\begin{table}
 \begin{tabular}{rlllllllll}
   $L_0$  & $q=2$            &  $q=3$           &  $q=4$
          & $q=8$            &  $q=8$           &  $q=16$
          & $q=16$           &  $q=64$          &  $q=64$          \\
          & $R=2$            &  $R=2$           &  $R=2$
          & $R=2$            &  $R=10$          &  $R=2$
          & $R=10$           &  $R=2$           &  $R=10$          \\ \hline
       1  &      0.563 (1)   &      0.927 (1)   &      1.184 (1)
          &      1.787 (1)   &      1.731 (3)   &      2.330 (1)
          &      2.322 (4)   &      3.120 (1)   &      3.476 (5)   \\
       2  &      0.508 (2)   &      0.825 (3)   &      1.042 (2)
          &      1.515 (3)   &      1.586 (10)  &      1.864 (3)
          &      2.101 (10)  &      2.194 (4)   &      3.150 (13)  \\
  {\bf 3} & {\bf 0.499 (3)}  & {\bf 0.800 (6)}  & {\bf 1.003 (6)}
          &      1.441 (7)   & {\bf 1.521 (23)} &      1.752 (8)
          & {\bf 2.052 (25)} &      2.065 (10)  & {\bf 3.034 (30)} \\
       4  &      0.500 (6)   &      0.813 (14)  &      0.994 (14)
          &      1.424 (15)  &      1.548 (52)  &      1.750 (17)
          &      2.089 (57)  &      2.157 (22)  &      3.079 (68)  \\
       5  &      0.505 (11)  &      0.842 (30)  &      1.005 (31)
          &      1.426 (33)  &      1.622 (113) &      1.785 (38)
          &      2.203 (125) &      2.305 (47)  &      3.209 (148) \\
       6  &      0.500 (20)  &      0.818 (62)  &      0.963 (63)
          &      1.360 (67)  &      1.587 (228) &      1.794 (78)
          &      2.196 (251) &      2.384 (93)  &      3.213 (300) \\
 \end{tabular}
 \caption{Effective central charge $c'$ extracted from parabolic fits with
   $L_0 \le L \le L_{\rm max}$, as described in the text. Error bars
   on the last quoted digit are shown in parentheses. The choice $L_0=3$
   appears to be optimal, provided that the strength of the randomness $R$
   is large enough (see text), and the corresponding values of $c'$, shown
   in bold face, should be regarded as our results.}
 \label{Tab:c}
\end{table}

\begin{table}
 \begin{tabular}{ccccccc}
  $p$ &$q=2$     &$q=4$     &$q=8$     &$q=16$    &$q=32$    &$q=64$ \\ \hline
  0.05&0.522 (25)&1.030 (26)&1.477 (27)&1.780 (29)&1.920 (30)&1.974 (31) \\
  0.10&0.519 (35)&1.032 (36)&1.510 (38)&1.915 (40)&2.251 (42)&2.552 (43) \\
  0.15&          &          &1.539 (46)&1.996 (48)&2.416 (50)&2.817 (52) \\
 \end{tabular}
 \caption{Effective central charge $c'$ obtained using a trinary
   distribution of the random bonds. There is a fraction $p$ of very weak
   and very strong bonds respectively, the remaining fraction $1-2p$ being
   critical.}
 \label{Tab:c-diluted}
\end{table}

\begin{table}
 \begin{tabular}{rllllll}
  $L_0$& $q=2$    & $q=3$    & $q=4$    & $q=8$    & $q=16$   & $q=64$    \\
       & $R=2$    & $R=2$    & $R=2$    & $R=10$   & $R=10$   & $R=10$    \\
  \hline
     2 &0.5662 (6)&0.9300 (7)&1.1903 (7)&1.723 (3) &2.313 (3) &3.469 (4)  \\
     3 &0.535 (1) &0.878 (1) &1.117 (1) &1.656 (5) &2.207 (6) &3.311 (7)  \\
     4 &0.521 (2) &0.848 (3) &1.075 (3) &1.605 (10)&2.148 (11)&3.206 (13) \\
     5 &0.514 (3) &0.836 (5) &1.051 (5) &1.585 (19)&2.125 (20)&3.163 (24) \\
     6 &0.512 (4) &0.839 (9) &1.043 (9) &1.595 (34)&2.144 (38)&3.174 (45) \\
     7 &0.510 (6) &0.840 (18)&1.022 (18)&          &          &           \\
     8 &0.509 (9) &0.812 (33)&1.011 (34)&          &          &           \\
     9 &0.503 (14)&          &          &          &          &           \\
    10 &0.500 (23)&          &          &          &          &           \\
 \end{tabular}
 \caption{Effective central charge $c'$ extracted from linear fits of
   $f_0^{11}(L) - f_0^{11}(\infty)$ versus $1/L^2$, with
   $L_0 \le L \le L_{\rm max}$. In all cases the approach towards the
   asymptotic values of Table III
% Should be \ref{Tab:c}, but RevTeX doesn't accept \ref macro here.
   is from above. Error bars
   on the last quoted digit are shown in parentheses.}
 \label{Tab:c-twopoint}
\end{table}

\begin{table}
 \begin{tabular}{c|rlrrrrr}
        & L & 1.~cumulant     & 2.~cumulant & 3.~cumulant & 4.~cumulant
                              & 5.~cumulant & 6.~cumulant \\ \hline
  $q=3$ & 1 & -1.039786 (242) &  0.060716   & -0.000791   &  0.000830
                              & -0.004583   & -0.000778   \\
        & 2 & -0.253209 (146) &  0.012391   & -0.000347   &  0.000279
                              &  0.000059   &  0.000386   \\
        & 3 & -0.106163 (113) &  0.004963   & -0.000246   &  0.000063
                              & -0.000102   &  0.000046   \\
        & 4 & -0.057901 (95)  &  0.002784   & -0.000143   &  0.000006
                              &  0.000034   &  0.000003   \\
        & 5 & -0.036521 (84)  &  0.001810   & -0.000105   &  0.000001
                              & -0.000002   & -0.000003   \\
        & 6 & -0.025172 (76)  &  0.001289   & -0.000075   &  0.000008
                              & -0.000001   & -0.000002   \\
        & 7 & -0.018426 (69)  &  0.000968   & -0.000069   &  0.000002
                              &  0.000002   & -0.000001   \\ \hline
  $q=8$ & 1 & -1.380171 (289) &  0.104382   &  0.004069   &  0.014001
                              &  0.019452   & -0.013889   \\
        & 2 & -0.326484 (177) &  0.028366   & -0.001683   & -0.000432
                              &  0.000157   & -0.003145   \\
        & 3 & -0.132560 (138) &  0.014908   & -0.001822   &  0.000356
                              &  0.000221   & -0.000083   \\
        & 4 & -0.071296 (115) &  0.010129   & -0.001610   &  0.000319
                              & -0.000959   &  0.002323   \\
        & 5 & -0.044886 (102) &  0.007880   & -0.001619   &  0.000252
                              & -0.000082   & -0.000456   \\
        & 6 & -0.031045 (92)  &  0.006450   & -0.001538   &  0.000607
                              & -0.000184   & -0.001096   \\
        & 7 & -0.022851 (84)  &  0.005401   & -0.001505   &  0.000237
                              &  0.000234   & -0.000280   \\
 \end{tabular}
 \caption{The first six cumulants of $-\Delta f(L)$ for $1 \le L \le 7$ and
   $R=2$. The error bar on the first cumulant (shown in parentheses) is
   related to the second cumulant; error bars on the higher cumulants are
   not shown.}
 \label{Tab:cum}
\end{table}

\begin{table}
 \begin{tabular}{rcc}
  $L$  & $\mu(L)$    & $\nu(L)$   \\ \hline
    1  &  1.087  (1) &  --        \\
    2  &  4.229  (3) &  1.041 (1) \\
    3  & 10.426  (8) &  0.816 (2) \\
    4  & 19.682 (18) &  0.827 (3) \\
    5  & 31.867 (33) &  0.863 (5) \\
    6  & 46.994 (53) &  0.885 (7) \\
    7  & 65.020 (79) &  0.904 (9) \\
 \end{tabular}
 \caption{Phenomenological renormalisation for the thermal scaling
   dimension $x_T = 2-1/\nu$ at $q=8$ and $R=10$. For each strip width $L$
   the 100 independent strips of length $m=10^5$ are divided into patches
   of length 200. Within each patch the {\em exact} $\mu(L,T_{\rm c})$ is
   computed, based on evaluations of $\xi(L,K_1',K_2')$ at 
   $K_1' = K_1 (1 \pm \epsilon_K)$ and $K_2' = R K_1'$, where $K_1$ is found
   from Eq.~(3).
% Should be Eq.~(\ref{selfdual}), but RevTeX doesn't accept \ref macro here.
   Final results and error bars are obtained as the mean value and the
   standard deviation over the totality of patches.}
 \label{Tab:phen-renorm}
\end{table}

\begin{table}
 \begin{tabular}{c|rcc}
         & $L$  & $\mu(L)$     & $\nu(L)$   \\ \hline
  $R=6$  &   2  &  1.898 (1)   & --         \\
         &   3  &  4.456 (2)   & 0.905 (1)  \\
         &   4  &  8.172 (6)   & 0.902 (2)  \\
         &   5  & 12.974 (11)  & 0.933 (4)  \\
         &   6  & 18.883 (18)  & 0.945 (6)  \\
         &   7  & 25.891 (27)  & 0.955 (8)  \\ \hline
  $R=10$ &   2  &  1.832 (1)   & --         \\
         &   3  &  3.917 (2)   & 1.144 (2)  \\
         &   4  &  6.815 (5)   & 1.081 (4)  \\
         &   5  & 10.486 (9)   & 1.074 (6)  \\
         &   6  & 14.948 (15)  & 1.059 (8)  \\
         &   7  & 20.198 (22)  & 1.050 (11) \\
 \end{tabular}
 \caption{Phenomenological renormalisation going perpendicularly off the
   critical surface for $q=8$ and $R=6$ and 10 respectively. The data
   collection was done as before.}
 \label{Tab:perpendicular}
\end{table}

\begin{table}
 \begin{tabular}{ccc}
  $q$ & $\nu$    & $R^{\ast}$ \\ \hline
    2 & 1.12 (3) &  7 (1)     \\
    3 & 1.04 (4) &  8 (1)     \\
    8 & 1.01 (2) &  9 (1)     \\
   64 & 1.02 (3) & 10 (1)     \\
 \end{tabular}
 \caption{Values of the critical exponent $\nu$ and the randomness strength
   $R^{\ast}$ at the random fixed point as obtained from phenomenological
   renormalisation.}
 \label{Tab:nu}
\end{table}

\begin{table}
 \begin{tabular}{clllll}
  $q$ & 1.~gap    & 2.~gap     & 3.~gap     & 4.~gap     & 5.~gap     \\ \hline
    3 & 0.899 (4) & 1.877 (13) & 1.885 (12) & 2.045 (24) & 2.050 (23) \\
    4 & 0.817 (5) & 1.811 (9)  & 1.818 (8)  & 2.043 (23) & 2.049 (24) \\
    5 & 0.754 (6) & 1.771 (6)  & 1.779 (6)  & 2.058 (24) & 2.065 (25) \\
 \end{tabular}
 \caption{Scaling dimensions corresponding to the first five gaps in the
   Lyapunov spectrum of ${\cal T}^{11}$ for $R=2$. The parabolic least-squares
   fits included the first three cumulants of the probability distribution,
   and error bars were extracted based on the fits with $L_0 = 4,5$ and 6.}
 \label{Tab:Lyapunov}
\end{table}

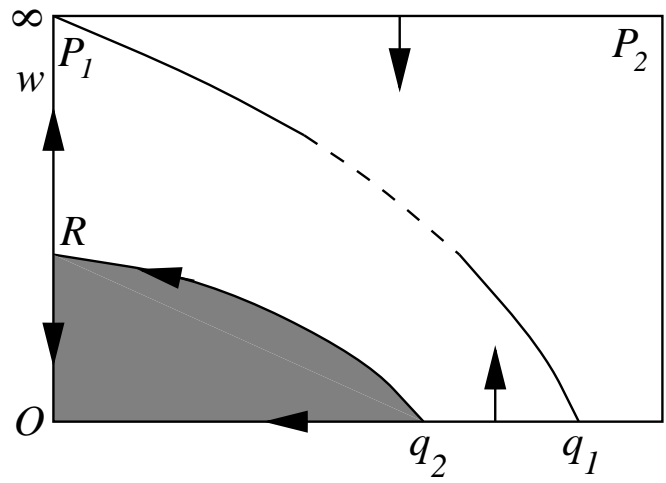
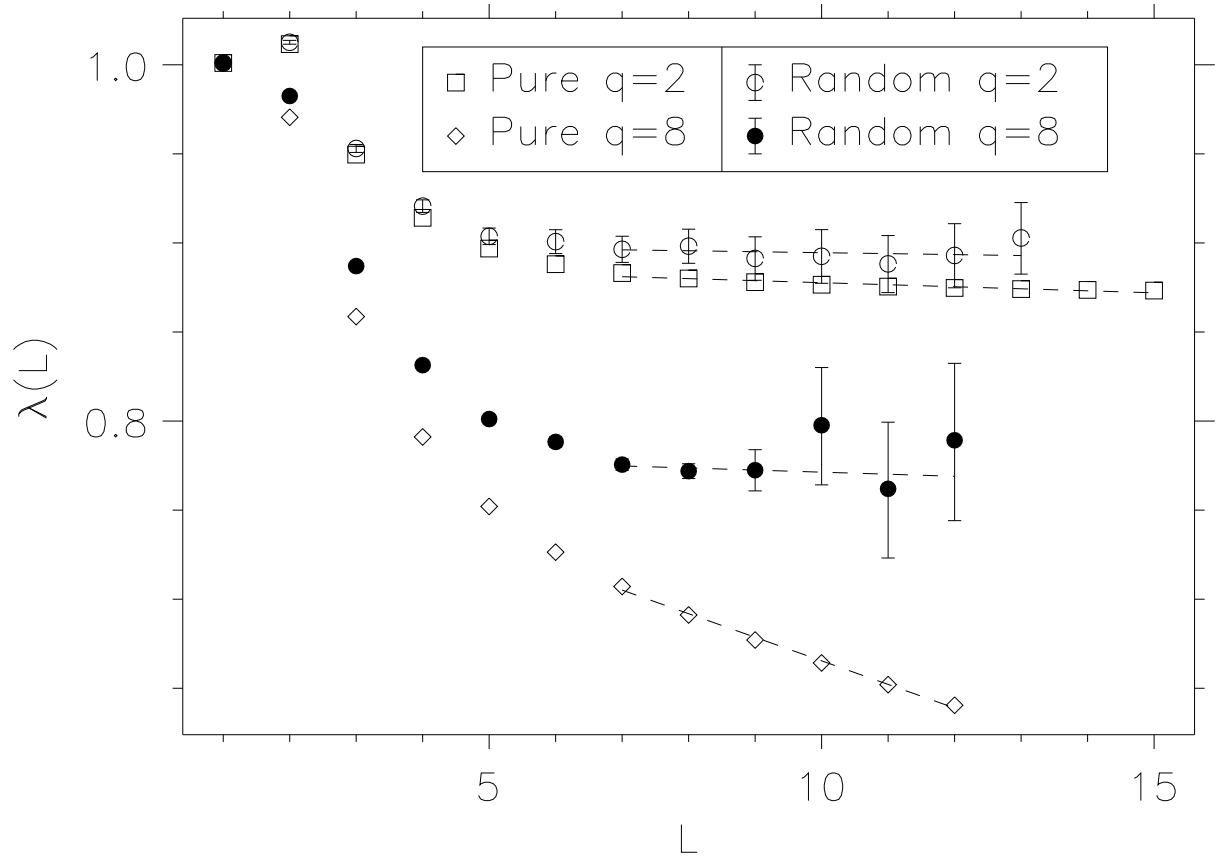
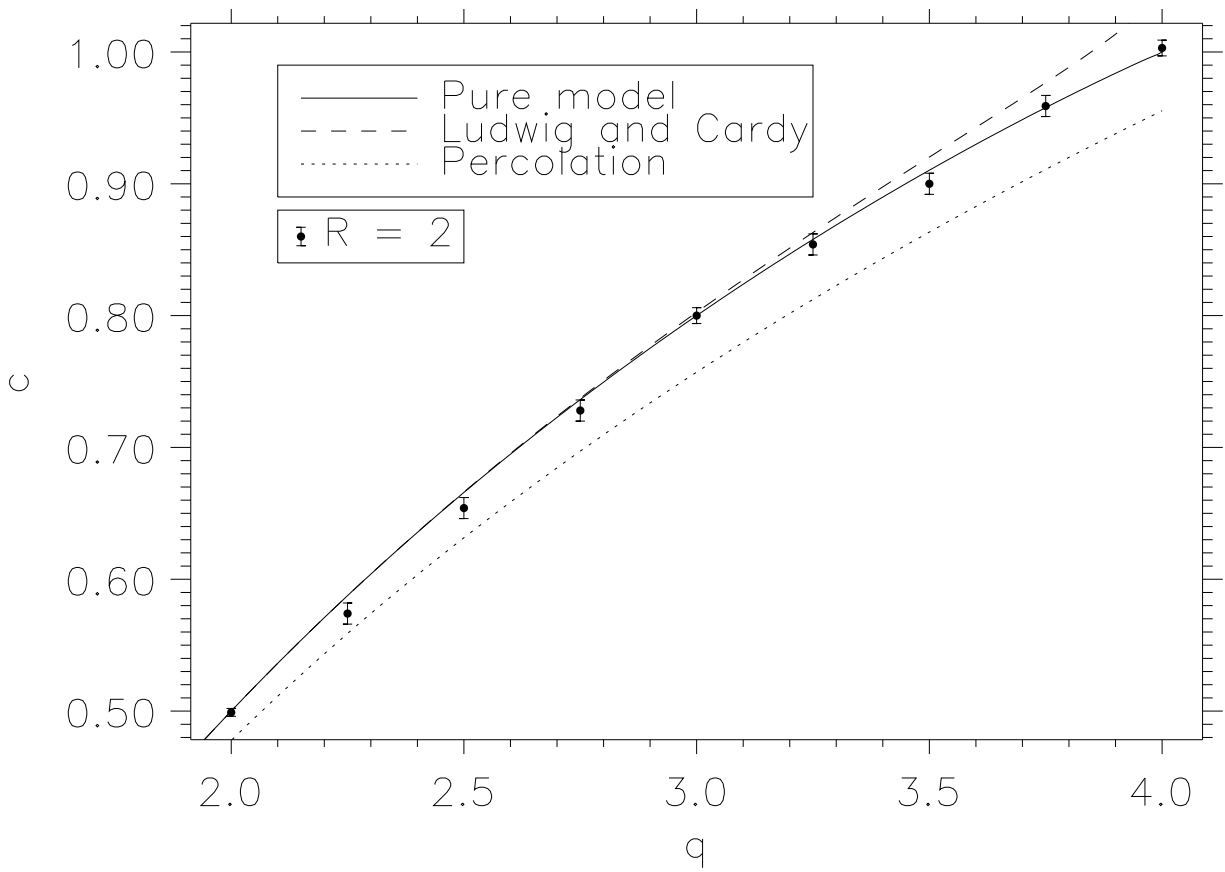
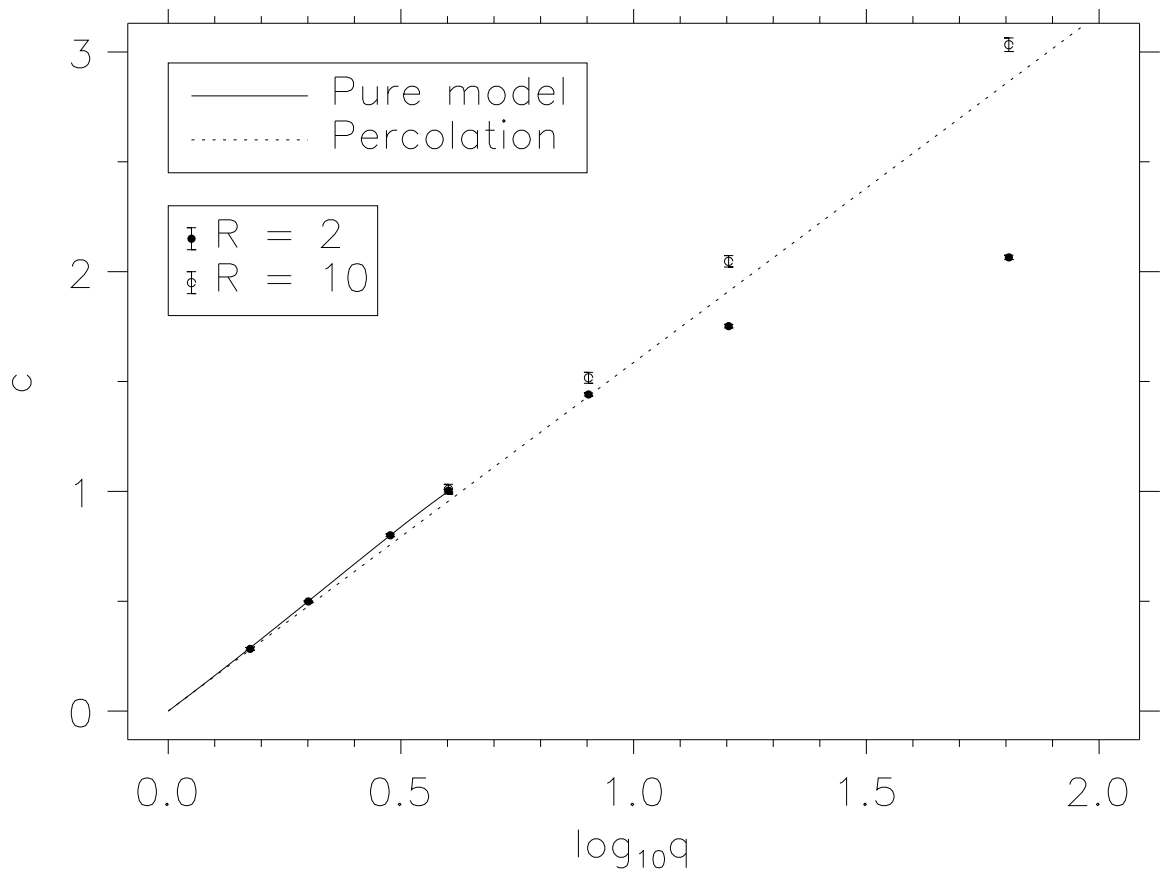
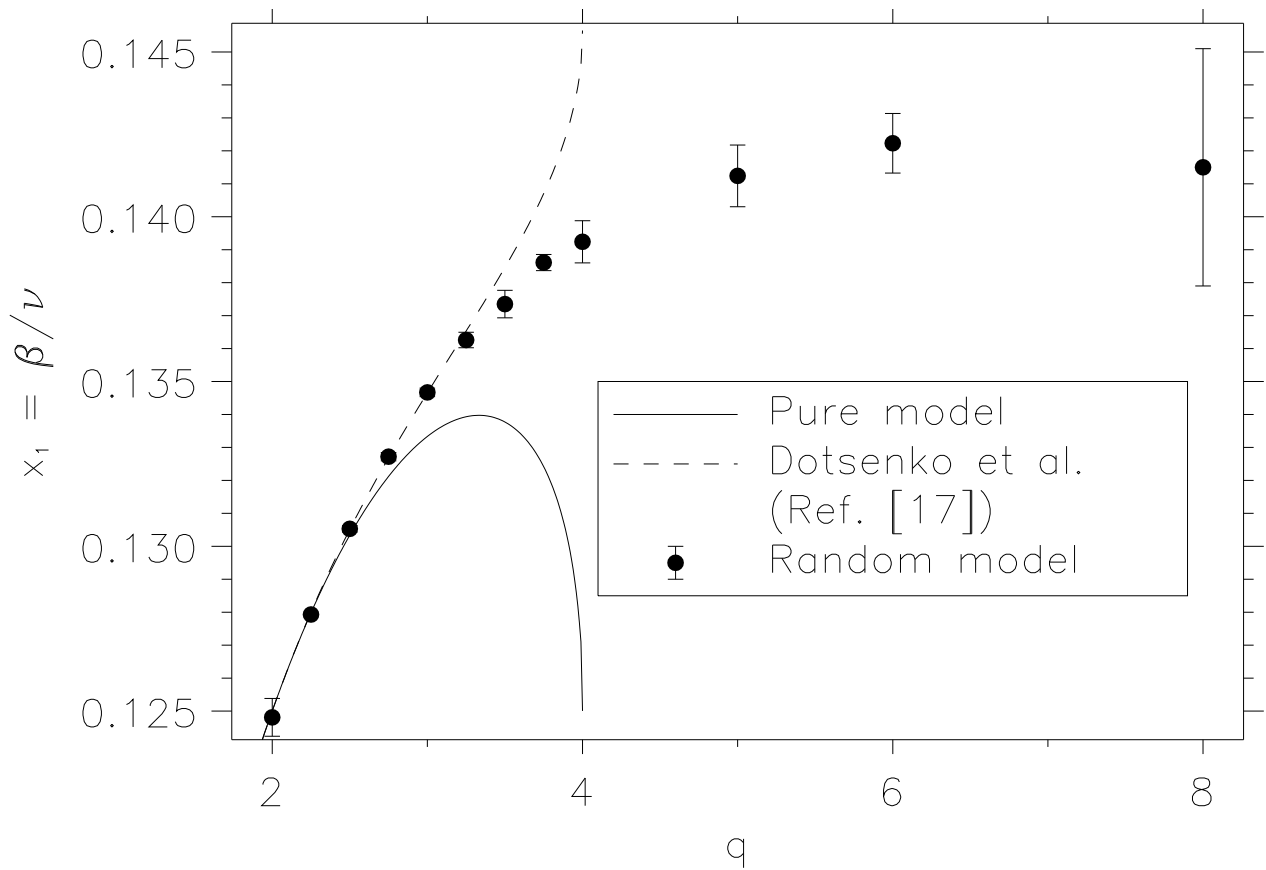
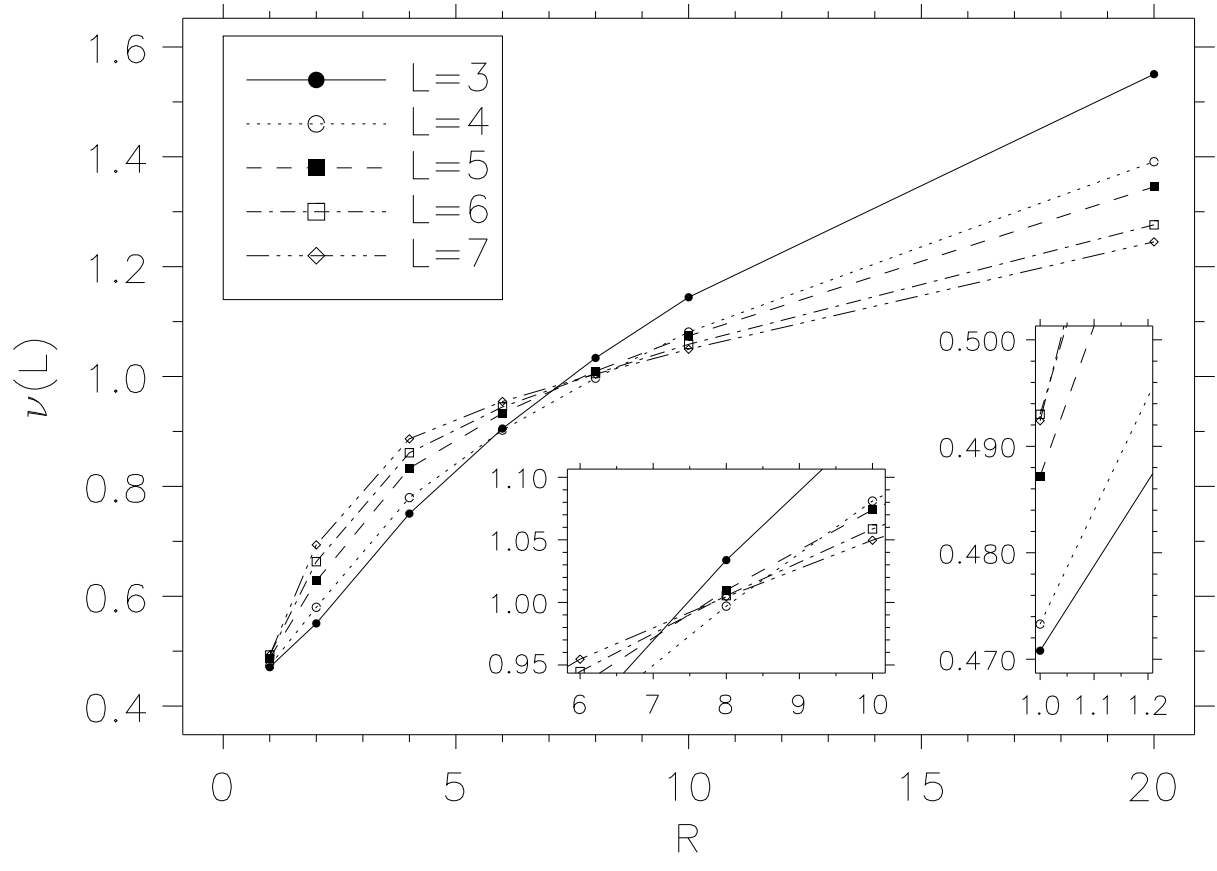
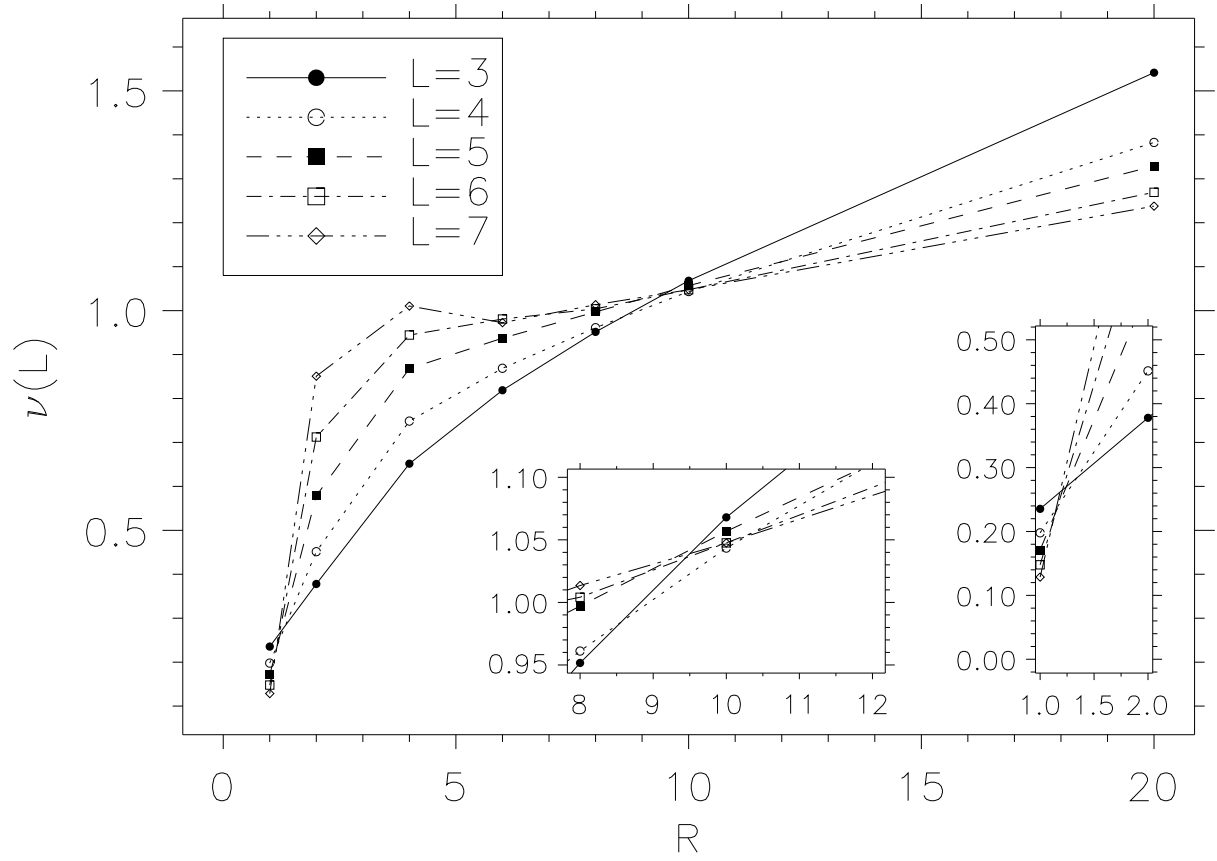
\end{document}